\documentclass[twocolumn,trackchanges]{aastex631}

\usepackage{float}
\usepackage{amsmath}
\usepackage{xcolor}
\usepackage{natbib}
\usepackage{longtable}
\definecolor{mycolor}{HTML}{bc3908} 
\definecolor{mycolor}{HTML}{d62828} 
\definecolor{mycolor}{HTML}{7b2cbf} 
\definecolor{mycolor1}{HTML}{0096c7} 
\definecolor{mycolor}{HTML}{bc4749}

\usepackage{footmisc}

\newcommand{\msun}{$\rm M_{\odot}$}

\renewcommand{\thefootnote}{\dag}

\begin{document}

% \title{
% %On the impact of intergalactic medium enrichment on 
% \yzq{Pre-enrichment of the intergalactic medium and C/O and N/O abundance ratios}
% %abundances 
% at $z>8$}

\title{{%The enrichment and 
On the importance of the IGM/CGM enrichment in determining chemical abundances during the first 600 Myr}}

\correspondingauthor{Annalisa Citro}
\email{acitro@shsu.edu, acitro@umn.edu}

\author[0009-0000-9676-0538]{Annalisa Citro}
%\thanks{acitro@umn.edu}
\affiliation{Department of Physics and Astronomy, Sam Houston State University, 1908 Avenue J, Huntsville, TX 77340, USA}
\affiliation{Minnesota Institute for Astrophysics, University of Minnesota, 116 Church Street SE, Minneapolis, MN 55455, USA}

\author[0000-0002-3146-2668]{
Yong-Zhong Qian}
\affiliation{Minnesota Institute for Astrophysics, University of Minnesota, 116 Church Street SE, Minneapolis, MN 55455, USA}

\author[0000-0002-9136-8876]{Claudia M. Scarlata}
\affiliation{Minnesota Institute for Astrophysics, University of Minnesota, 116 Church Street SE, Minneapolis, MN 55455, USA}

% \author[0000-0002-9946-4731]{Marc Rafelski}
% \affiliation{Space Telescope Science Institute, 3700 San Martin Drive, Baltimore, MD 21218, USA}
% \affiliation{Department of Physics and Astronomy, Johns Hopkins University, Baltimore, MD 21218, USA}

% \author[0000-0002-4917-7873]{Mitchell Revalski}
% %\altaffiliation{AASTeX v6+ programmer}
% \affiliation{Space Telescope Science Institute, 3700 San Martin Drive, Baltimore, MD 21218, USA}



\begin{abstract}
{JWST observations show that galaxies at $z > 8$ exhibit a wide range of carbon-to-oxygen (C/O) ratios at fixed metallicity and systematically elevated nitrogen-to-oxygen (N/O) ratios compared to local galaxies. Proposed explanations include finely tuned star-formation histories (SFHs), differential or pristine inflows, top- or bottom-heavy initial mass functions, or enrichment from Wolf–Rayet stars, supermassive stars (SMSs), or Population~III stars. The observed abundances reflect both in situ production tied to the SFHs and earlier pre-enrichment of the intergalactic and circumgalactic medium (IGM/CGM), which supply the gas inflow that fuels star formation. At such early epochs, SFHs are necessarily brief, and sustained star formation requires inflows to dominate over outflows, making the chemical state of the IGM/CGM especially important. We develop a chemical-evolution model that incorporates (1) enrichment of the IGM/CGM by multiple sources and (2) the observed SFHs of high-redshift galaxies. Applying this model to eight galaxies at $z > 8$, we find that only specific pre-enrichment scenarios reproduce their observed abundance ratios. For example, the low C/O and high N/O ratios measured in GN-z11 and CEERS-1019 require an IGM/CGM enriched by SMSs and pre-enriched to an oxygen metallicity of  $\sim10^{-5}$ of the solar value. We also find that the $\log(\mathrm{N/O})$ vs. $12 + \log(\mathrm{O/H})$ plane provides more stringent constraints on enrichment pathways than the $\log(\mathrm{C/O})$ vs. $12 + \log(\mathrm{O/H})$ plane. Overall, our results highlight the crucial role of IGM/CGM chemical composition in chemical-evolution models at $z > 8$ and underscore the significant influence of SMSs on early carbon, nitrogen, and oxygen abundances.}\\

\noindent
\textit{\href{https://astrothesaurus.org/concept-select/}{Unified Astronomy Thesaurus concepts}}: Abundance ratios (11), Early universe (435), Galaxy chemical evolution (580), Chemical enrichment (225)

\end{abstract}



\section{Introduction} \label{sec:intro}

The baryon cycle is central to a galaxy's evolutionary path. Gas is accreted from the {intergalactic medium (IGM) and circumgalactic medium (CGM)}, transformed into stars, processed within them, released back into the interstellar medium (ISM) through stellar winds or explosions, and partly ejected back into the CGM/IGM through outflows \citep[see review by][]{MaiolinoMannucci2019}. A crucial aspect of the baryon cycle is that different elements are synthesized by stars of varying mass and are released into the ISM on different timescales. Therefore, relative abundance ratios between different chemical elements are key to understanding the galaxy life cycle, as they provide information about the types of stars that produce them, as well as the properties of the gas flowing into and out of the galaxy.

Among chemical elements,
%abundances, the absolute oxygen abundance (O/H) and the relative ratios involving 
carbon, nitrogen, and oxygen (C, N, and O)
%(such as C/O, N/O, and O/H) 
are particularly significant. This is not only because they
%oxygen, carbon, and nitrogen 
are the most abundant elements in the Universe after hydrogen and helium, but also because they trace galaxy evolution. Oxygen is a primary element, i.e., it is synthesized from hydrogen and helium without requiring any
%the presence of 
pre-existing heavier elements. O is produced in massive stars (with  $\rm M \gtrsim 8$ \msun) and is mainly released into the ISM by core-collapse supernovae (CCSNe) on timescales shorter than 100 Myr. Carbon, also a primary element {(but see, for example, \citealp{Henry+2000, Berg+2019b, Berg+2021, Chiappini+2003a} for discussion of possible secondary production of C)}, is produced by any stars with $\rm M \gtrsim 1$ \msun\ and is mainly released into the ISM through the stellar winds of asymptotic giant branch (AGB) stars on timescales longer than 100 Myr (for reference, an intermediate mass star of $\sim6$ \msun\ lives $\sim 110$ Myr on the
%has a long 
main sequence). 
%lifetime). 
Nitrogen behaves as both a primary and a secondary element (as a secondary element, its synthesis depends on 
%the presence of 
pre-existing C and O
in the stellar interior), and the two synthesis paths of N have contributions from both massive and low- to intermediate-mass stars \citep[see review by][]{Romano2022}. 
%Since oxygen 
{As a crude approximation, the O/H abundance can serve as a proxy for evolutionary time (effects of gas inflows and outflows can complicate the relation between O/H and time). Moreover, as O is exclusively released by massive stars, while C and N are also released by lower-mass stars, C/O and N/O ratios provide insights into the SFHs involving different stellar populations.
This characteristic makes O/H, C/O, and N/O excellent observables for reconstructing the chemical evolution path of galaxies. In addition, chemical evolution of the earliest galaxies is expected to be dominated by massive stars, and extraordinary C/O and N/O ratios would suggest distinct sources even among massive stars. Such possibilities are now testable with the James Webb Space Telescope (JWST) enabling observations of the earliest galaxies.} 

Studies of C/O and N/O as a function of O/H began with observations of Milky Way stars \citep[e.g.,][]{Gustafsson+1999, Fabbian+2009} and \ion{H}{2} regions \citep[e.g.,][]{KobulnickySkillman1996, Afflerbach+1997, Mattsson2010}. These studies revealed that C/O and N/O follow a flat trend at 12 + log(O/H) $ < 8$, with low averages of $\rm log(C/O)\sim -0.7$ and $\rm log(N/O)\sim -1.6$, respectively, while increasing with gas metallicity for 12 + log(O/H) $> 8$. The flat trends at low O/H are attributed to the primary production of C and N exclusively by massive stars at early times, while the metallicity dependent behaviour at higher O/H is usually explained by
%with 
either delayed/secondary contributions to
%production of 
C and N from intermediate-mass stars, or  metallicity-dependent stellar winds of massive stars \citep{Chiappini+2003b, Romano+2020, Romano2022}.

Research on C/O and N/O vs. O/H has been subsequently extended to galaxies both in the near and far Universe {\citep[e.g.,][]{Berg+2019b, Jones+2023, ArellanoCordova+2022, Stiavelli+2023, MarquesChaves2024, Curti+2025, Carniani+2024, Citro+2024, Deugenio+2024, Hsiao+2024}}. However, because the observed abundances in galaxies are the results of multiple generations of stars, additional factors must be taken into account, such as the galaxy age, the star formation efficiency, and the duration and separation of the star-forming bursts. Moreover, as
%since 
galaxies are not closed systems, the observed abundances also depend on the inflow and outflow of heavy elements \citep[e.g.,][]{Vincenzo+2016, Berg+2019b}. This is true especially for galaxies with
%at 
low masses and metallicities, which
%where galaxies 
have shallow potential wells \citep[e.g.,][]{DekelSilk1986, Dalcanton2007, PeeplesShankar2011}.

%abundance 
{In the local Universe, $\log\text{(C/O)}$ exhibits significant scatter ranging from $\sim -1$ to $\sim -0.3$ within the metallicity interval $7 \lesssim 12 + \log(\mathrm{O}/\mathrm{H}) \lesssim 8$ (\citealp{Tsamis+2003, Esteban+2004, Esteban+2009, Esteban+2014, GarciaRojas+2004, GarciaRojas+2005, GarciaRojas+2007, Peimbert+2005, LopezSanchez+2007, Berg+2016, Berg+2019b, ToribioSanCipriano+2016, ToribioSanCipriano+2017, PenaGuerrero+2017, Senchyna+2017, Ravindranath+2020, Senchyna+2021}). A similar degree of scatter has also been detected %in the C/O 
in galaxies at higher redshifts, including those at $z \gtrsim 8$ (when the Universe was less than 650 Myr old) \citep{Berg+2019b, Jones+2023, ArellanoCordova+2022, Stiavelli+2023, MarquesChaves2024, Curti+2025, Carniani+2024, Citro+2024, Deugenio+2024, Hsiao+2024}, as well as in stacked galaxy samples at $z \sim 4$--$7$ \citep{Hu+2024, Hayes+2025}.} 
%However, there are exceptions. For instance, \citet{Hsiao+2024} and \citet{Deugenio+2024} reported high $\rm log(C/O)\gtrsim-0.44$ at low oxygen metallicities ($\rm 12+log(O/H) <7.8$). 
{At $z > 8$, the observed scatter in the C/O abundance ratio has been attributed to the evolutionary stage of the galaxy: the {more evolved the system is}, the higher its C/O ratio, as more intermediate- and low-mass stars have had time to evolve off the main sequence and contribute carbon through the AGB phase \citep[e.g.,][]{Jones+2023, Curti+2025, Hsiao+2024}. Another possible explanation involves differential galactic outflows {(i.e., outflows that remove CCSNe products more efficiently than AGB products, see \citealp[e.g.,][]{Rizzuti+2024})}. The more oxygen-rich the expelled material {is}, the higher the resulting C/O ratio in the ISM \citep{Berg+2019b}. Additionally, extremely high C/O abundances at $z\sim12$ have been attributed to more exotic stellar populations, such as Population III stars \citep{Deugenio+2024}.}

{Concerning nitrogen, galaxies at $z > 8$ systematically show N/O ratios
%abundances 
that significantly deviate, up to a factor of $\sim40$, from those in the local Universe \citep[e.g.,][]{Isobe+2023,MarquesChaves2024, Topping+2025a, Topping+2025b}.}
{These high N/O ratios are comparable to those observed in nearby globular clusters, suggesting that high‑redshift systems with similar chemical signatures may represent their progenitors \citep[e.g.,][]{Senchyna+2024}.}
Several explanations have been proposed to account for such high N/O, including fine tuned SFHs \citep{KobayashiFerrara2024}, differential outflows, top heavy or bottom heavy initial mass functions (IMFs, \citealp{Curti+2025, ArellanoCordova+2024}), {enrichment by Wolf-Rayet stars \citep{KobayashiFerrara2024, Berg+2025}}, pristine inflows diluting all abundances but
not affecting abundance ratio O and leaving the C/O and N/O ratios unchanged \citep{Stiavelli+2025, Morishita+2025a}, Very Massive Stars \citep{MarquesChaves+2026}, Wolf-Reyet stars \citep[e.g.,][]{Gunawardhana+2025} or AGNs \citep[e.g.,][]{Zhu+2026}. Another compelling possibility involves enrichment by more exotic stellar populations, such as Population III (Pop III) stars or Supermassive stars (SMS) \citep[e.g.,][]{Nagele+2023a, Charbonnel+2023, MarquesChaves2024, Ebihara+2026}.

{Pop~III stars are primordial, initially metal-free stars thought to form at $z \sim 15$--$20$ with masses between $\sim 10$ and $260\,M_{\odot}$ \citep[e.g., see the review by][]{BrommLarson2004, KlessenGlover2023}. Although born metal-free, they synthesize C, N, and O internally through helium burning (triple-$\alpha$), subsequent $\alpha$-capture, and the CNO cycle once a small amount of carbon is present. Their explosion properties strongly influence their chemical yields. Low-energy Pop~III supernovae (POPIII-LE hereafter), with $E_{\rm released} \sim 0.3\times10^{51}$~ergs, eject only the outer stellar layers, leading to substantial fallback of iron-group and oxygen-rich material onto the remnant and producing ejecta enhanced in carbon and nitrogen. In contrast, high-energy Pop~III explosions (POPIII-HE hereafter), with $E_{\rm released}\sim10^{52}$~ergs, eject also the deeper, oxygen-rich layers of the star and therefore, yield much larger amounts of oxygen \citep{HegerWoosley2010, Chen+2017, Vanni+2023}.}

{SMSs are predicted to form over a wide metallicity range, from $Z = 0$ to $Z \sim Z_{\odot}$ \citep{Nagele+2023a}, through (at least) two distinct channels. In one channel, they assemble in low-mass halos of $M_{\mathrm{halo}} \sim 10^7$–$10^8\,M_{\odot}$ via runaway stellar collisions in very dense clusters \citep[e.g.,][]{Chen+2014, Woods+2021, Nagele+2022}. Alternatively, they can form in gas-rich mergers between massive halos, either by direct gravitational collapse of primordial gas, or by runaway stellar collisions following gas fragmentation at higher metallicity, typically on Myr timescales \citep[e.g.,][]{Mayer+2010,MayerBonoli2019,Nagele+2023a}. In both scenarios, an SMS with mass $M_{\star} \sim 10^3$–$10^5\,M_{\odot}$ can form. Such a star ultimately becomes unstable to general-relativistic (GR) collapse; if significant nuclear fuel (e.g., hydrogen during core H-burning through the CNO cycle) remains at the onset of GR instability, late-time explosive burning can bring the star to explode. In models where the SMS is disrupted instead of collapsing directly to a black hole, this explosive phase ejects large quantities of nitrogen-rich material, consistent with the extreme N/O ratios observed in some high-redshift systems \citep[e.g.,][]{Nagele+2023a}.}

{Chemical evolution models are one of the primary tools used to interpret the chemical abundances of galaxies across a wide range of redshifts}.
{Despite the significant progress achieved with such modeling in understanding how factors such as SFHs \citep[e.g.,][]{KobayashiFerrara2024}, stellar types \citep[e.g.,][]{Nagele+2023a}, IMFs \citep[][]{ArellanoCordova+2025b}, and outflow properties \citep[e.g.,][]{Rizzuti+2024} affect the chemical abundances at high redshift ($z\gtrsim5$), the properties and initial chemical composition of the inflowing gas have received minimal attention. Nevertheless, the inflowing gas accreted from the IGM and CGM fuels star formation, with feedback subsequently returning some of the processed material back into these reservoirs. Simulations show that the CGM attains a chemical composition similar to that of the galaxy ISM, since it is enriched primarily by galactic outflows, which can extend well into the CGM (up to $\sim200$~kpc from the galaxy) \citep{CenChisari+2011, Shen+2012, Shen+2013}. Observational and theoretical studies collectively show that the IGM metal abundance increases toward higher gas overdensities \citep[e.g.,][]{Schaye+2003, Oppenheimer+2009}, and that its metallicity spans $10^{-4}$--$10^{-1}\,Z_{\odot}$ \citep{Oppenheimer+2009, DOdorico+2016, DiStefano+2026}.} {Because the metallicity of the inflowing gas provides the baseline enrichment of the ISM, the chemical composition of the IGM/CGM can potentially influence the evolutionary pathways of galaxies.}

{Instead of sophisticated simulations, we present a model based on simple analytical prescriptions for chemical evolution of galaxies at high redshift. In the framework of hierarchical structure formation, the dark matter halo hosting a galaxy grows through accreting surrounding small halos or merging with a relatively large halo. During the earlier phase, the potential well of the host halo is shallow and feedback from star formation in the halo drives strong outflows that feed the surrounding CGM/IGM. Such outflows repeatedly disrupt the accumulation of gas that is supplied by inflows into the halo. Consequently, the total mass of stars formed during this phase is insignificant compared to what is achieved later. The net result for chemical evolution during this phase can be approximated as setting the metallicity and composition of the CGM/IGM, which we refer to as the ``pre-enrichment'' and treat only parametrically but not in detail. Specifically, we assign a level of pre-enrichment and consider various massive stellar sources (standard massive stars, Pop~III stars, and SMSs) to fix the composition. The pre-enriched CGM/IGM provides the inflows for the subsequent phase, during which the potential well of the host halo is so deep that inflows dominate outflows and the total gas and stellar masses grow steadily. This later phase is the focus of our model, which depends on the pre-enrichment, the inflow and outflow rates, the SFH and associated stellar populations, and the nucleosynthesis yields for the pertinent stellar sources.}

{While our model is simple and parametric, it is in qualitative agreement with cosmological simulations. Galaxies of interest to us have $z\gtrsim 8.5$ and total stellar masses of $\sim 10^9\,M_{\odot}$ or less. They resemble the halo D9 in \cite{Bassini+2023}, whose Figs. 1, 4, and 5 clearly show that the overall evolution of D9 is divided into an earlier outflow-dominated phase followed by an inflow-dominated one. In fact, this two-phase evolution also extends to the larger halos D7 and D3 (see Fig. 1 in \citealt{Bassini+2023}), except that the duration of their outflow-dominated phase is shorter. Based on the above discussion, we consider that our model is reasonable for treating chemical evolution of those high-redshift galaxies of interest. A key ingredient of our model is pre-enrichment of the CGM/IGM providing the inflows that dominate the later phase of chemical evolution.}

This paper is organized as follows: in Section \ref{sec:data} we describe the literature data used in this work, specifically focusing on the SFHs; in Section \ref{CEM} we describe our chemical evolution model and summarize the implemented analytical equations; %in Section \ref{sec:IGM-pre}, \ref{sec:ism_enr}, and \ref{sec:out} we discuss the main ingredients used in our chemical evolution model; 
in Section \ref{sec:res} we illustrate the results; in Section \ref{sec:discussion} we present our discussion; our conclusions are summarized in Section \ref{sec:conclusions}. {Additional plots and further discussion of the results are presented in Appendix \ref{sec:appendix_model}.} 
Throughout the paper, we assume a Plank18 \citep{Planck2018} cosmology, and a solar composition characterized by $\rm 12 + log(O/H)_{\odot} = 8.69$, $\rm log(C/O)_{\odot}=-0.26$, and $\rm log(N/O)_{\odot}=-0.86$ \citep{Asplund+2009}.

\section{Data}
\label{sec:data}

\begin{figure*}[htbp]
    \centering

    % Left panel
    \includegraphics[width=0.49\textwidth]{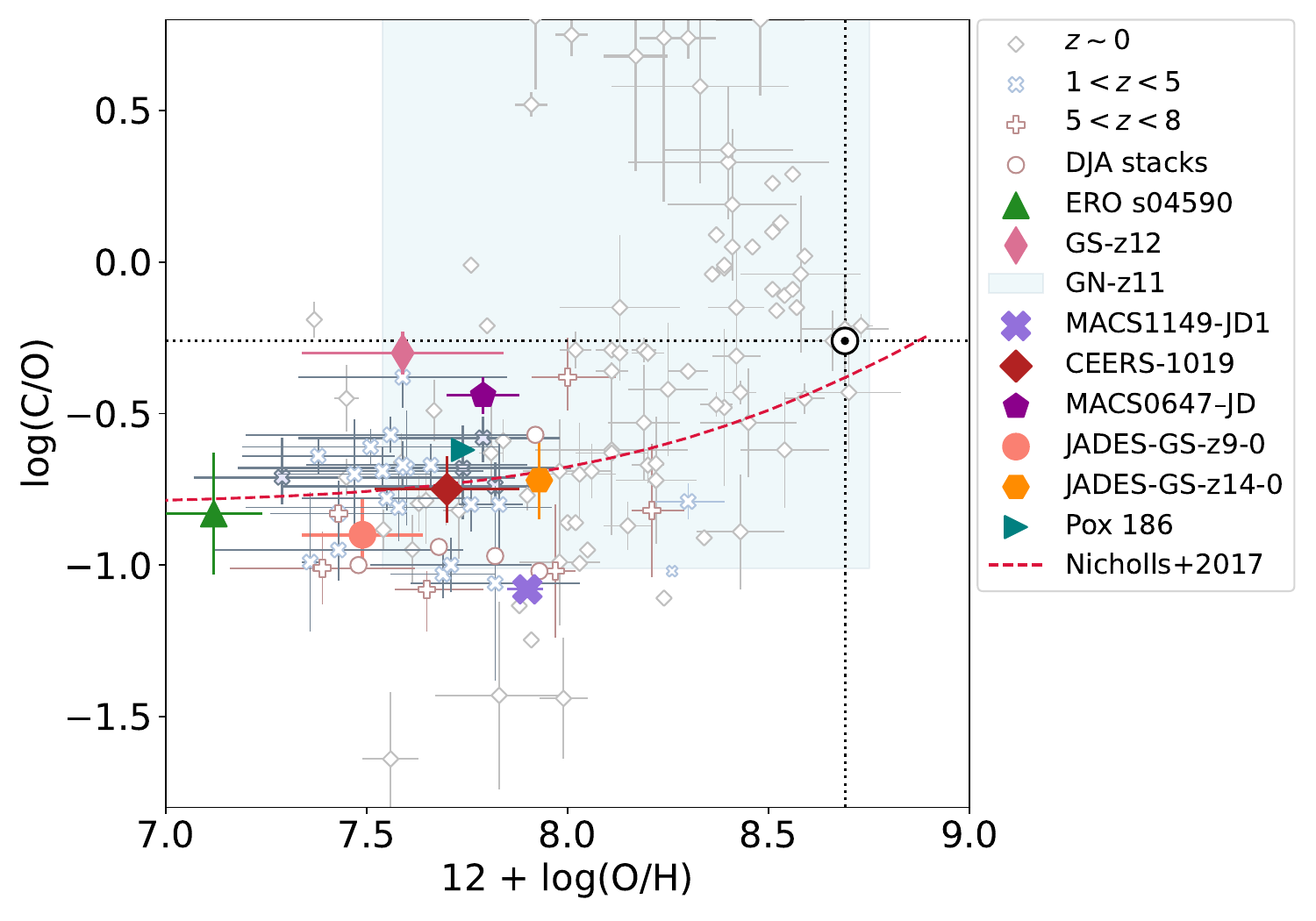}
    \hfill
    % Right panel
    \includegraphics[width=0.49\textwidth]{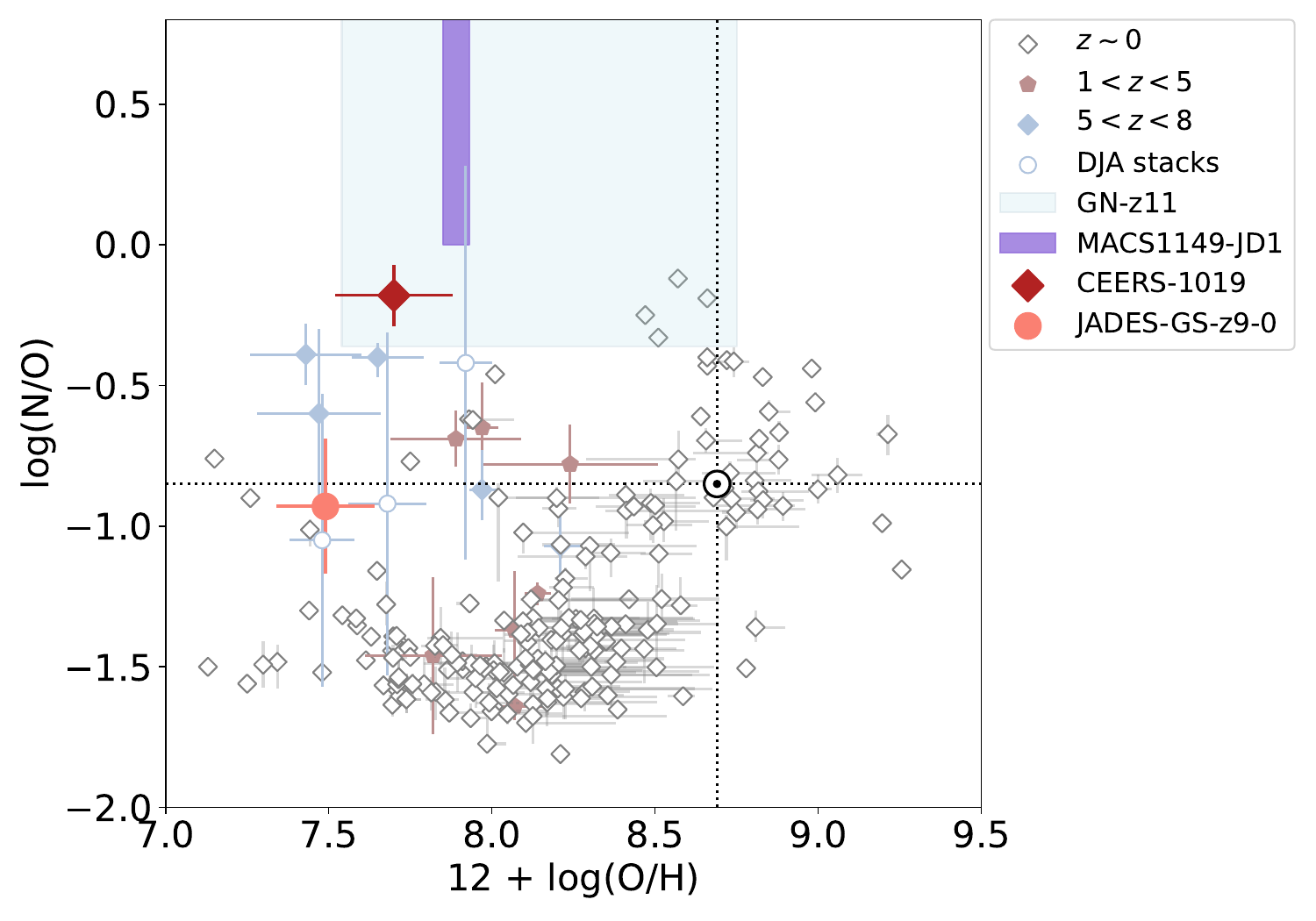}

\caption{log(C/O) vs. 12 + log(O/H) (left) and log(N/O) vs. 12 + log(O/H) (right) planes for galaxies at different redshift. 
{Left panel}: Results at $z\sim0$ are shown as small white diamonds \citep{Tsamis+2003, Esteban+2004, Esteban+2009, Esteban+2014, GarciaRojas+2004, GarciaRojas+2005, GarciaRojas+2007, Peimbert+2005, LopezSanchez+2007, Berg+2016, Berg+2019b, ToribioSanCipriano+2016, ToribioSanCipriano+2017, PenaGuerrero+2017, Senchyna+2017, Ravindranath+2020, Senchyna+2021}. Results at $z\sim1$--$5$ are shown as light blue {times symbols} \citep{Erb+2010, Christensen+2012a, Bayliss+2014, James+2014, Stark+2014, Steidel+2016, Amorin+2017, Mainali+2020, Matthee+2021, Iani+2023}. %Damped Ly$\alpha$ systems (DLAs) at $z\sim1$--$4$ are shown as small upward triangles \citep{Cooke+2017, Saccardi+2023a}.
Values for galaxies at $5<z<8$ are shown as {plus symbols} and are taken from \citet{Jones+2023,Isobe+2023,Topping+2024a,Topping+2024b,Ji+2024}.  {Stacked spectral data from \citet{Hayes+2025} are shown as small circles}. Bigger colored symbols mark our sample galaxies \citep{ArellanoCordova+2022, MarquesChaves2024, Stiavelli+2023, Curti+2025, Hsiao+2024, Isobe+2023, Deugenio+2024, Carniani+2024}. Note that GN-z11 has only a lower limit on log(C/O), shown as a shaded rectangular area. The red dashed curve is the fit from \citet{Nicholls+2017}; the black dotted lines and the sun symbol mark the solar values for log(C/O) and 12 + log(O/H).
{Right panel}: Results at $z\sim0$ are shown as white diamonds \citep{Berg+2016, Berg+2019b, ValeAsari+2016}; results for $1<z<5$ are shown as small brown pentagons \citep{Christensen+2012a, James+2014, Bayliss+2014, Stark+2014, Steidel+2016, Berg+2018, Matthee+2021, Sanders+2023b, Rogers+2024, Welch+2024}; results at $5<z<8$ are shown as light blue diamonds \citep{Isobe+2023, Topping+2024a, Topping+2024b, ArellanoCordova+2025b}. We also include the stacked spectral data from \citealp{Hayes+2025} as white circles. N/O ratios for galaxies at $z>8$ are shown as big colored symbols \citep{MarquesChaves2024, Curti+2025}; note that GN-z11 and MACS1149-JD1 have only lower limits on log(N/O), shown as shaded rectangular areas.
}
\label{fig:logfig}
\end{figure*}

We select eight $z>8$ literature targets with $\rm 12+log(O/H)$ and C/O measurements, including four with N/O estimates. The selected galaxies are ERO s04590 \citep{ArellanoCordova+2022}, CEERS-1019 \citep{MarquesChaves2024}, MACS1149-JD1 \citep{Stiavelli+2023}, JADES-GS-z9-0 \citep{Curti+2025}, MACS0647-JD \citep{Hsiao+2024}, GN-z11 \citep{Cameron+2023, Isobe+2023}, GS-z12 \citep{Deugenio+2024}, and JADES-GS-z14-0 \citep{Carniani+2024}. {The abundance estimates, together with other relevant galaxy properties (redshift, stellar mass, SFR) are listed in Table~\ref{tab:highzCO}. These galaxies are also indicated on the $\rm 12+log(O/H)$ vs. $\rm log(C/O)$ and $\rm log(N/O)$ planes shown in Figure~\ref{fig:logfig}.}

\begin{figure*}[!ht]
\centering
     \includegraphics[width = 1\textwidth]%{plot_targets_SFH_3_new.pdf}
     {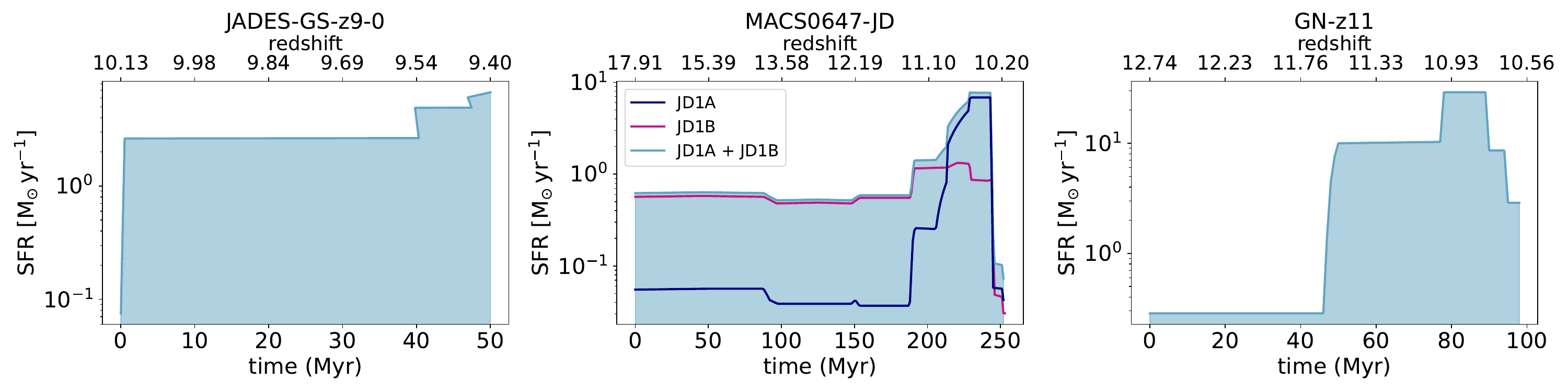}
    \caption{Photometry-derived SFHs for selected galaxies in our sample. We show the photometry-derived SFHs for JADES-GS-z9-0 \citep{Curti+2025}, MACS0647–JD \citep{Hsiao+2023}, and GN-z11 \citep{Tacchella+2023}. For MACS0647–JD, {\citet{Hsiao+2023} identified two emission knots, named JD1A and JD1B. For these two emission knots, they derived two different individual SFHs, which we show in the figure, together with the the combined SFH, which we use in our chemical evolution model}. For ERO s04590, we adopt the photometry-derived SFH from \citet{Carnall+2023}, which is extremely short ($<3$ Myr) and therefore not shown. For galaxies with only stellar mass and SFR estimates available, we assume a constant SFH extending over the time required to build up the observed stellar mass; these galaxies are also not shown.}
\label{fig:sfh}
\end{figure*}

{In the following Sections, we will apply our novel chemical evolution model (described in Section \ref{CEM}) to each galaxy using their photometry‑derived SFH, shown in  Figure \ref{fig:sfh} for JADES-GS-z9-0 \citep{Curti+2025}, MACS0647–JD \citep{Hsiao+2023}, and GN-z11 \citep{Tacchella+2023}. For MACS0647-JD, we show the SFHs for the two {emission knots JD1A and JD1B identified by \citet{Hsiao+2023}} along with the combined SFH that we adopt in our chemical‑evolution model.} For ERO s04590, we assume the SFH derived by \citet{Carnall+2023}, which is extremely short ($<3$ Myr) and not shown.

If the SFH is not available, but {estimates of the galaxy's stellar mass ($M_{\rm\star, obs}$) and star formation rate (${\rm SFR}_{\rm obs}$)} are, 
we assume a constant SFH over the time period: 

\begin{equation}
t_{\rm SFH} = M_{\rm \star,obs}/{\rm SFR}_{\rm obs}.~~
\end{equation}

\begin{deluxetable*}{rcccccccc}[ht]
\tabletypesize{\footnotesize}
\tablecolumns{9} 
\tablecaption{\label{tab:highzCO} {Properties of the galaxies} investigated in this work}%, and their O/H, C/O and N/O abundances. } 
\tablehead{\colhead{ID} & \colhead{$z\,^a$} & \colhead{${\rm log(M_\star/M_{\odot})}$} & \colhead{SFR$_{\rm obs}\,^b$ [$\rm M_{\odot}\,yr^{-1}$]} & \colhead{12+log(O/H)} & \colhead{log(C/O)} &  \colhead{log(N/O)}  &  \colhead{SFH$\,^c$} &  \colhead{References$\,^d$}}
\startdata
ERO s04590 & 8.49 & $7.10\substack{+0.14 \\ -0.12}$ & 4.21 ($\sim 3$ Myr) & $7.12 \pm 0.12$  & $-0.83 \pm 0.38$  & -- & NP  & (1)\\
CEERS-1019 &  8.68 & $9.30\substack{+0.13 \\ -0.13}$ & $161\pm23$ (10 Myr) & $7.70 \pm 0.18$ & $-0.75 \pm 0.11$  & $-0.18 \pm0.11$ & C  &  (2)\\
MACS1149-JD1 & 9.11 & $8.20 \substack{+0.06 \\ -0.06}$ & $5.9\pm0.2$& $7.90\,^{+0.04}_{-0.05}$ &  $-1.08 \pm 0.04$   & $< 0.00$ & C  &  (3)\\
JADES-GS-z9-0 & 9.40 & $8.17 \substack{+0.02 \\ -0.02}$ & $4.34 \substack{+0.10 \\ -0.08}$ (10 Myr) & $7.49 \pm 0.11$  & $-0.90 \pm 0.12$  & $-0.93 \pm 0.24$  &  NP  &  (4)\\
 &  & &  &  &  & $< -0.60$  &   & \\
MACS0647–JD & 10.17 & $8.62 \substack{+0.15 \\ -0.11}$ & $4\pm1$ (100 Myr) &  $7.79 \pm 0.09$   &  $-0.44\,^{+0.06}_{-0.07}$ & -- & NP  & (5)\\
GN-z11 & 10.60  &$9.10 \substack{+0.40 \\ -0.30}$& $21.00\substack{+0.22 \\ -0.10}$ (30 Myr) &  $8.00\,^{+0.46}_{-0.76}$ & $> -1.01$ & $> -0.36$ & NP  & (6)\\
GS-z12 &  12.50 & $7.68\substack{+0.19 \\ -0.19}$ & $1.62\pm0.01$&  $7.59 \pm 0.25$ &  $-0.30 \pm 0.07$    & -- & C & (7)\\
%&  &- &- &  $<6.9\,^c$ &  $>-0.21\,^c$   & - &  & \\ 
%&  &- & -&  $<6.9\,^d$ &  $>-0.36\,^d$   &- &   & \\ 
JADES-GS-z14-0 & 14.0 & $8.29\substack{+0.14 \\ -0.09}$ & $14.45\substack{+0.07 \\ -0.10}$ & $7.95 \pm 0.2$  &  $-0.72 \pm 0.13$ & -- & C  & (8)\\
\enddata 
{\tablenotetext{$\tiny$ a}{Redshift.}
\vspace{-2mm}
\tablenotetext{$\tiny$ b}{An entry with an interval in the parentheses indicates that SFR$_{\rm obs}$ is the average SFR over this interval.}
\vspace{-2mm}
\tablenotetext{$\tiny$ c}{NP means a non-parametric (photometry-derived) SFH and C means an assumed constant SFR.}
\vspace{-2mm}
\tablenotetext{$\tiny$ d}{(1) \citet{ArellanoCordova+2022,Carnall+2023}; (2) \citet{MarquesChaves2024}; (3) \citet{Stiavelli+2023}; (4) \citet{Curti+2025}; (5) \citet{Hsiao+2023,Hsiao+2024}; (6) \citet{Tacchella+2023,Isobe+2023}; (7) \citet{Deugenio+2024}; (8) \citet{Carniani+2024}. For JADES-GS-z9-0, \citet{Curti+2025} provided a measurement of N/O based on the detection of the N IV]$\lambda$1483 line, and an upper limit from the 3$\sigma$ nondetection of the N III]$\lambda\lambda$1747–1754 muliplet. For GN-z11, emission lines were measured by \citet{Maiolino+2024} and \citet{Bunker+2023}. For GS-z12, the abundance ratios were derived from the \texttt{BEAGLE} SED fitting.}
}
\end{deluxetable*} 

\section{Chemical evolution model}
\label{CEM}

{As discussed in Section \ref{sec:intro}, we treat the overall evolution of a galaxy in two distinct phases. In the earlier phase, {the host halo of the galaxy does not have a sufficiently deep potential well and star formation drives strong outflows that disrupt accumulation of gas in the halo. We assume that this complex history results in a pre-enriched IGM/CGM through ejection of metals by those outflows. This pre‑enriched IGM/CGM supplies the inflowing gas that fuels subsequent star formation in the galaxy’s later evolutionary phase, during which a typical, smoother gas–star–gas cycle becomes established. By “smoother” we mean a regime in which outflows no longer repeatedly disrupt the accumulation of gas delivered by inflows; instead, feedback operates at a level that allows the galaxy to retain a significant fraction of the incoming gas and to process it continuously. In this phase, inflows, star formation, and outflows can be approximately described by continuous functions instead of occurring in short, highly sporadic bursts.} Our model focuses on the build-up of C, N, and O in the galaxy's ISM during the later phase, taking into account {pre-enriched} inflows, outflows, and the galaxy's specific SFH. The use of the SFH inferred from data (see Section \ref{sec:data}) removes the uncertainties related to the timescales of heavy element release in the ISM, leaving only the stellar type, inflow, and outflow to constrain. 

The starting point (time $t=0$) of our model coincides with the onset of the galaxy’s observed SFH.}
{We assume that all the gas in the later phase is provided by the inflow with a rate}
%In our chemical evolution model we assume that a galaxy is characterized by an initial stellar mass $M_{s,0}=0$ and  initial gas mass is $M_{g,0} = 0$. We define the inflow rate $f_{inf}(t)$ as:
\begin{equation}
\label{eq:inflow}
%f_{inf}(t) = \frac{M_{\rm inf}}{\tau_{\rm inf} \times (1 - e^{-t_{obs}/\tau_{\rm inf}})} \times e^{(t/\tau_{\rm inf})}
f_{\rm inf}(t) = \frac{M_{\rm inf}\times e^{t/\tau_{\rm inf}}}{\tau_{\rm inf} \times (e^{t_{\rm obs}/\tau_{\rm inf}}-1)},
\end{equation}
where $M_{\rm inf}$ is the \textit{total} accreted gas mass, $t_{\rm obs}$ is the time {at observation (numerically equal to the duration of the SFH)}, and $\tau_{\rm inf}$ is the inflow timescale (reflecting how quickly the inflowing gas is accreted).

%\vspace{0.2cm}
%\noindent
The outflow rate %$f_{out}(t)$ 
is defined as:
\begin{equation}
f_{\rm out}(t) = w \times \psi(t),
\label{eq:outflow}
\end{equation}
where $\psi(t)$ is the SFR at time $t$, and $w$ is {a constant outflow efficiency} parameter.% that we assume.\\

%\noindent
We model the time evolution of the gas mass $M_{g}$ {[with $M_g(0)=0$]} as:

\begin{equation}
   \dot{M}_{g}(t) = f_{\rm inf}(t) - f_{\rm out}(t) + e(t) - \psi(t)\ \, , 
\end{equation}
%where $f_{inf}(t)$ is the inflow rate, $f_{out}(t)$ is the outflow rate, and $\psi(t)$ is the SFR at time $t$. 
%\vspace{0.2cm}
%\noindent
where $e(t)$ {is the rate at which gas is ejected} into the ISM by all stars, and is defined as:

\begin{equation}
  e(t) = \int_{m(t)}^{m_{\rm up}}{(m - r(m))\, \psi(t - \tau(m))\, \phi(m)\,dm}.
  \label{eq:mg}
\end{equation}
In this equation:
\begin{itemize}
\item[-] $m$ is the mass of the star {in units of \msun};
\item[-] $m(t)$ is the mass of the star {that was born at $t=0$ and is} exiting the main sequence at time $t$;
\item[-] $m_{\rm up}$ is the upper mass limit;
\item[-] $r(m)$ is the remnant mass left behind; 
\item[-] $\tau(m)$ is the main sequence lifetime of a star of mass $m$ {and $\tau(m(t))=t$};
\item[-] $t - \tau(m)$ is the {birth time for a star of mass $m$ that is exiting the main sequence at time $t$};
\item[-] $\psi(t - \tau(m))$ is the SFR at the time $t-\tau(m)$;
\item[-] $\phi(m)$ is the IMF, {normalized as $\int_{m_{\rm low}}^{m_{\rm up}}m\phi(m)dm=1$ with $m_{\rm low}$ being the lower mass limit}.
\end{itemize}

We model the time evolution of the stellar mass $M_s$ {[with $M_s(0)=0$]} as: 
\begin{equation}
   \dot{M_{s}}(t) = \psi(t) - eS(t)\, ,
\end{equation}
{where $eS(t)$ {is the rate of loss to stars that are dying at time $t$}}:
\begin{equation}
eS(t) = \int_{m(t)}^{m_{\rm up}}{m\, \psi(t - \tau(m))\, \phi(m)\,dm}.
\end{equation}

%\noindent

%\vspace{2mm}
The time evolution of the mass $M_{\rm X}$ of element X (${\rm X=C}$, N, O) in the ISM (gas) is defined as:

\begin{equation}
\begin{split}
%\dot{M}_{\rm X}(t) = e_{\rm X}(t) - Z_{g,{\rm X}}(t) \times \psi(t) + Z_{\rm inf, X} \times f_{\rm inf}(t)\\
%-  Z_{g, X}  \times q_X \times f_{out}(t)
\dot{M}_{\rm X}(t) = e_{\rm X}(t) - Z_{g,{\rm X}}(t) \times \psi(t) + Z_{\rm inf,X}\times f_{\rm inf}(t)\\
-  Z_{g,{\rm X}}(t)  \times q_{\rm out,X} \times f_{\rm out}(t).
\label{eqZ}
\end{split}
\end{equation}
%\noindent
In this equation,
\begin{itemize}

%item $e(t)_{X}$ is the mass of element X that is ejected by stars into the ISM
\item[-] {$Z_{g,{\rm X}}(t)=M_{\rm X}(t)/M_g(t)$ is the mass fraction of element X in the gas};
\item[-] {$Z_{\rm inf,X}$ is the constant mass fraction of element X in the inflow};
%$Z_{\rm inf,O}$ is varied between $10^{-6}$ and $10^{-4}$};
%\item[-] $Y_{X/O}$ is the ratio between the stellar yield of element $X$ ($Y_{X}$) and the oxygen yield ;
%\item[-] $f_{inf}(t)$ is the inflow rate;
\item[-] $q_{\rm out,X}$ is the {preferential factor for loss of element $X$} through outflows;
%\item[-] $f_{out}(t)$ is the outflow rate;
\item[-] $e_{\rm X}(t)$ is {the rate at which element $X$ is ejected into the gas} by stars, and
similarly to Eq. (\ref{eq:mg}), %$e_{X}(t)_{X}$ is computed as:
% \begin{equation}
% e_{\rm X}(t)= \int_{m(t)}^{m_{\rm up}}{Y_{\rm X,tot}(m,Z_{Y}(t)) \times\frac{\psi(t-\tau(m))}{M_{\odot}} \times \phi(m)\, dm },
% \label{eq:etx}
% %\end{split}
% \end{equation}
\begin{equation}
\begin{split}
e_{\rm X}(t)= 
\int_{m(t)}^{m_{\rm up}}
\mathbf{Y_{\rm X,tot}(m,Z_g(t-\tau(m)))} \times\ \\
\frac{\psi(t-\tau(m))}{M_{\odot}} \times\, \phi(m)\, dm
\end{split}
\label{eq:etx}
\end{equation}

where
    %\item[-]  $(m - r(m)) \times Z_{g,X}(t-\tau(m))$ is the mass of element $X$ that a star of mass $m$ removed from the ISM when it was formed and is now being released back to the ISM through stellar winds;
{$Y_{\rm X,tot}(m,Z_g(t-\tau(m)))$ is the total mass of element X ejected at its death by a star of mass $m$ formed at time $t-\tau(m)$, including both the mass already present at its birth and the newly formed mass $Y_{\rm X,new}$. {The dependence of $Y_{\rm X,tot}$ on the initial metallicity $Z_g(t-\tau(m))$ of the star takes into account both the mass of X corresponding to this metallicity and the dependence of $Y_{\rm X,new}$ on this metallicity.}} 
\end{itemize}

\begin{figure*}[htbp]
\includegraphics[width = \textwidth]%{plot_yields_by_mass_apr26.pdf}
{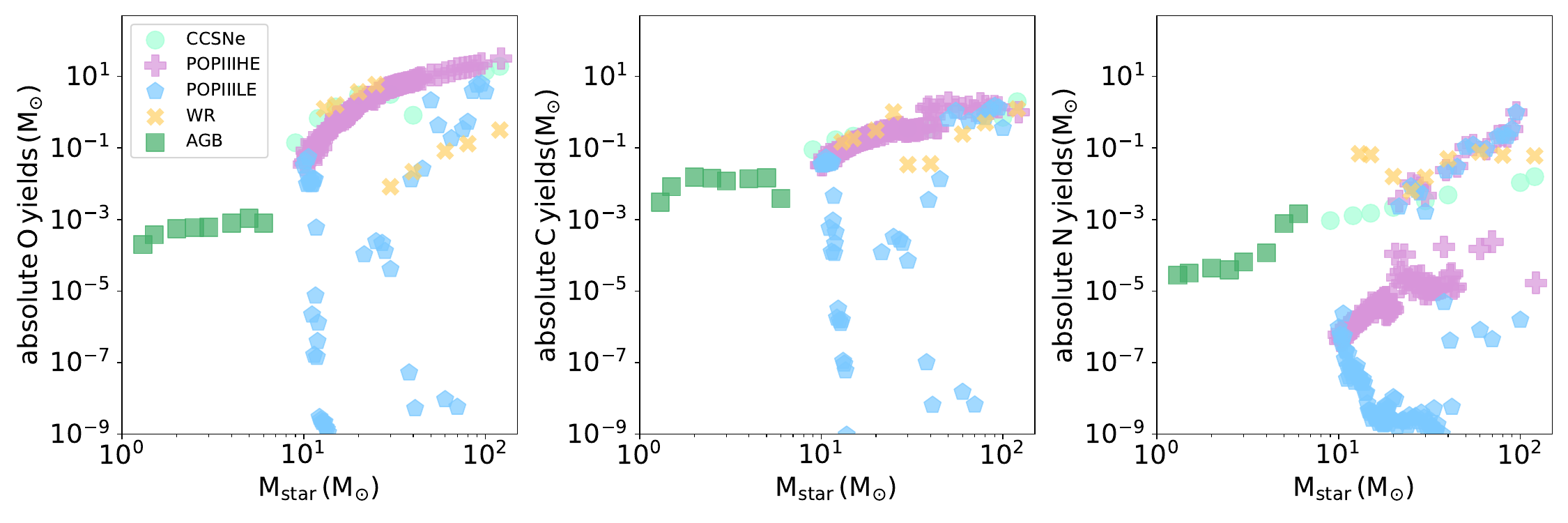}
\caption{{Total mass yields of oxygen, carbon and nitrogen as a function of stellar mass adopted in our chemical evolution model. CCSN yields are taken from \citet{Portinari+1998}; POPIII yields are taken from \citet{HegerWoosley2010}; AGB yields are taken from the F.R.U.I.T.Y database \citep{Cristallo+2016}. Wolf-Rayet yields (taken from \citealp{LimongiChieffi2018}) are shown for completeness even though they are not directly used in our model. It is evident that adjacent progenitor masses can show significant jumps in the O and C yields for low‑energy Pop III explosions, and in the N yields for both low‑ and high‑energy Pop III explosions. These discontinuities likely arise from abrupt changes in the stellar shell structure, compactness, or carbon‑burning behavior across neighboring masses. }}
\label{fig:yields}
\end{figure*}

{With $\psi(t)$ for $0\leq t\leq t_{\rm obs}$ taken from the SFH inferred from observations (see Section~\ref{sec:data}), the above system of equations can be solved once we specify
\begin{itemize}
\item[-]$M_{\rm inf}$, $\tau_{\rm inf}$, $Z_{\rm inf,X}$ for the inflow;
\item[-]$w$ and $q_{\rm out,X}$ for the outflow;
\item[-]$m_{\rm low}$, $m_{\rm up}$, $\phi(m)$, $\tau(m)$, and $Y_{\rm X,tot}(m,Z_g)$ for the stars.
\end{itemize}
Before discussing the assumptions made for these quantities (in Sections \ref{sec:IGM-pre}, \ref{sec:ism_enr}, and \ref{sec:out}), we note that our main focus is in exploring the inflow parameters while keeping the other quantities fixed. We are interested in the effects on O/H, C/O, and N/O at the age of the galaxy, which can be obtained from $({\rm X/H})_{\rm obs}=Z_{g,{\rm X}}(t_{\rm obs})/A_{\rm X}Z_{g,{\rm H}}$. Here $A_{\rm X}$ is the atomic mass number of element X and $Z_{g,{\rm H}}=0.75$ is taken to be the fixed mass fraction of hydrogen in the gas. As expected from the above system of equations and confirmed by the results to be presented in Section~\ref{sec:dependencies}, the (O/H)$_{\rm obs}$ is sensitive to $M_{\rm inf}$ and $\tau_{\rm inf}$ for the inflow, while the corresponding (C/O)$_{\rm obs}$ and (N/O)$_{\rm obs}$ are sensitive to $Z_{\rm inf,C}/Z_{\rm inf,O}$ and $Z_{\rm inf,N}/Z_{\rm inf,O}$, respectively. The last two quantities are determined by the sources that pre-enriched the IGM/CGM}, {and the relative importance of the pre-enrichment will depend on the duration of the SFH. {We highlight that our model only prescribes the form of the inflow rate and the different sources for the pre-enrichment of the inflow, but does not make any specific assumptions on the total mass or characteristic timescale for the inflow.}}

\subsection{Inflowing gas}
\label{sec:IGM-pre}

{We choose $M_{\rm inf}$ and $\tau_{\rm inf}$ to ensure that
%such that, at any time, the integrated total gas mass is higher than the integrated total stellar mass. In this way, we ensure that, at any time, 
enough gas is present to sustain star formation at any time. We vary $Z_{\rm inf,O}$ (see below) and take $Z_{\rm inf,X}=Z_{\rm inf,O}\times(Y_{\rm inf,X}/Y_{\rm inf,O})$, where $Y_{\rm inf,X}$ and $Y_{\rm inf,O}$ are the effective yields of element X and O for the stellar sources that have pre-enriched the inflow. We assume four such sources: standard massive stars that evolve as CCSNe, Population III stars with low ($0.3\times10^{51}$~ergs, POPIII-LE) and high explosion energy ($10^{51}$~ergs, POPIII-HE), and a single SMS with $M = 10^{5}$ \msun}.

For CCSNe, we use yields from \citet{Portinari+1998} for stars {with masses between 9 and 120 $\rm M_{\odot}$ and {a metallicity of $0.02\, Z_{\odot}$}.} {Compared to other yields \citep[e.g.,][]{LimongiChieffi2018}, the \citet{Portinari+1998} yields also account for the mass loss due to Pre-CCSN winds}. For POPIII-LE and POPIII-HE, we use yields from \citet{HegerWoosley2010} {for stars with masses between 10 and 120 $\rm M_{\odot}$.} {These yields are particularly suitable because they provide a clear separation in explosion energies for Pop III stars, allowing us to test both the low‑energy and high‑energy scenarios in a consistent way. {We adopt the \(10^{5}\,{M}_{\odot}\) SMS} yields from \citet{Nagele+2023a}, which are consistent with those used in other studies assessing the presence of SMSs in high redshift galaxies \citep[e.g.,][]{MarquesChaves2024} and therefore allow a consistent comparison.} 
The effective yields of the above sources are obtained as 
\begin{align}
    Y_{\rm inf,X}=\frac{\int_{m_1}^{m_2}Y_{\rm X}(m)m^{-\alpha}dm}{\int_{m_1}^{m_2}m^{-\alpha}dm},
\end{align}
where $Y_{\rm X}(m)$ is the yield of element X as a function of stellar mass $m$, $m_1$ and $m_2$ specify the relevant mass interval, and $\alpha$ is the power index of the assumed power-law IMF. Due to the absence of metals and the correspondingly inefficient cooling in the earliest star-forming environments -- which promotes the formation of more massive stars (e.g., \citealp{StacyBromm2013, Hirano+2014, Jaura+2022, Prole+2022a}; see also the review by \citealp{KlessenGlover2023}) -- we adopt a relatively top-heavy IMF with a slope of $\alpha = 2.1$. The IMF-integrated effective yields of C, N, and O for CCSNe, POPIII-HE, and POPIII-LE are summarized in Table~\ref{tab:poll}. For comparison, we also give the results for $\alpha = 1, 2.35$, and $3$. For the SMS case, we list the yields of a single SMS of mass $10^5$ {\msun}.

{Figure~\ref{fig:yields} shows how the C, O, and N yields vary with stellar mass and stellar type. The yields are single-valued for a specific mass although adjacent progenitor masses can show large jumps in the O and C yields (especially for POPIII-LE), and N yields (for both POPIII-LE and POPIII-HE). As discussed by \citet{HegerWoosley2010}, small changes in initial mass lead to different shell structures, compactness, and carbon‑burning behavior, which in turn modify shock propagation and fallback even under identical explosion conditions. This naturally produces substantial yield variations from one mass to the next. However, our prescribed pre-enrichment only depends on the IMF-averaged yields listed in Table~\ref{tab:poll}.}

{We use $Z_{\rm inf,O}$ to parameterize the absolute level of IGM/CGM pre-enrichment and adopt distinct values depending on the sources of pre-enrichment. For POPIII-HE and POPIII-LE, we assume $Z_{\rm{inf,O}} = 10^{-6}$, {which, assuming a solar pattern, corresponds to a total metallicity $Z_{\text{inf}} \sim 3\times10^{-6}$. This value reflects the primordial conditions under which POPIII stars form. For CCSNe, we assume $Z_{\text{inf,O}} = 10^{-6}, 10^{-5},\ 5 \times 10^{-5},\ 10^{-4}$, {which corresponds to the total $Z_{\text{inf}} = 3\times10^{-6}, 3\times 10^{-5}, 5 \times 10^{-4},\ 3\times 10^{-4}$}. For SMS, we assume $Z_{\text{inf,O}} = 10^{-5},\ 5 \times 10^{-5},\ 10^{-4},\ 10^{-3}$, {which corresponds to $Z_{\text{inf}} = 3\times 10^{-5}, 10^{-4},\ 3\times 10^{-4},3\times 10^{-3}$. Our range of $Z_{\rm{inf}}$ spans the simulated IGM metallicities at $z\sim5-8$ across different levels of gas overdensity \citep{Oppenheimer+2009}.}}

\subsection{ISM enrichment}
\label{sec:ism_enr}

{The inflowing gas described in the previous section triggers star formation}, which we model using the SFH inferred from observations (see Section~\ref{sec:data} and Figure~\ref{fig:sfh}). For each SFH, we assume a power-law IMF $\phi(m)\propto m^{-\alpha}$ with a range of  $\alpha = 1$, 1.5, 2.1, 2.35, 2.5, 2.8, 3.0, 3.2. For massive stars (9--$120\,M_\odot$), we adopt the yields from \citet{Portinari+1998}, which cover the metallicity range of 0.02--$2.5\,Z_{\odot}$ (the yields at the lowest metallicity are also employed for modeling IGM/CGM pre-enrichment). For low- and intermediate-mass stars (1--6 $M_{\odot}$), we use the yields from the F.R.U.I.T.Y. database \citep{Cristallo+2009, Cristallo+2011, Cristallo+2016}. These stars live rather long lives on the main sequence ($\sim 110$ Myr for a $\sim6 \,M_{\odot}$ star), and release C, N, and O into the ISM during the subsequent AGB phase (note that the galaxies considered in this work have SFH durations of $\lesssim$ 250 Myr, so AGB stars have minimal impact on the predicted abundances). We note that Eq.~(\ref{eq:etx}) is consistent with our adopted yields from \citet{Portinari+1998} and \citet{Cristallo+2016}, which are defined as $Y_{\rm X,tot}(m,Z_g) = (m - r(m))M_\odot \times Z_{g,{\rm X}}(t_{\rm birth}) + Y_{\rm X,new}$.

\subsection{Outflowing gas}
\label{sec:out}

{We assume that the outflow rate is proportional to the SFR (see Eq.~(\ref{eq:outflow})) and consider three values of the efficiency parameter $w = 5$, 100, 500. Following similar assumptions made for galaxies at both low \citep{Berg+2019b} and high redshift \citep{Rizzuti+2024}, we adopt a preferential factor $q_{\rm out,X}=0.3$, 0.3, and 1 for loss of C, N, and O through outflows, respectively.}

\begin{deluxetable*}{cccc}[ht]
\tabletypesize{\footnotesize}
\tablecolumns{4} 
\tablecaption{\label{tab:poll} {Effective yields of C, N, and O for sources pre-enriching the IGM/CGM}}
\tablehead{\colhead{Source} & \colhead{$\log(Y_{\rm inf,C}/\mbox{\msun})\,^a$} & \colhead{$\log(Y_{\rm inf,N}/\mbox{\msun})\,^a$} &  \colhead{$\log(Y_{\rm inf,O}/\mbox{\msun})\,^a$}}
\startdata
\multicolumn{4}{c}{{$\alpha = 2.1$}}\\
\vspace{-2mm}
CCSNe & $-0.59$ & $-2.59$ & $~~0.26$\\
POPIII-HE & $-0.43$  & $-1.74$ & $~~0.57$\\ 
POPIII-LE & $-1.07$ & $-1.75$ & $-0.65$\\
\hline
\multicolumn{4}{c}{{$\alpha = 2.35$}}\\
\vspace{-2mm}
CCSNe & $-0.63$ & $-2.64$ & $~~0.21$\\
POPIII-HE & $-0.50$  & $-1.88$ & $~~0.48$\\ 
POPIII-LE & $-1.19$ & $-1.89$ & $-0.80$\\
\hline
\multicolumn{4}{c}{{$\alpha = 3$}}\\
\vspace{-2mm}
CCSNe & $-0.71$ & $-2.75$ & $~~0.08$\\
POPIII-HE & $-0.68$  & $-2.27$ & $~0.23$\\ 
POPIII-LE & $-1.48$ & $-2.28$ & $-1.20$\\
\hline
\multicolumn{4}{c}{{$\alpha = 1$}}\\
\vspace{-2mm}
CCSNe & $-0.40$ & $-2.34$ & $~~0.59$\\
POPIII-HE & $-0.15$  & $-1.24$ & $~~0.96$\\ 
POPIII-LE & $-0.61$ & $-1.24$ & $-0.10$\\
\hline
SMSs &  $~~0.75$ & $~~1.91$ & $-0.05$\\
\enddata 
%\rm log(C/O)_{IGM}$ & -0.75 &  -0.21  & -0.82 & -0.17\\
%\tablenotetext{$\tiny$*}{$<Y_X>_{\rm IGM} = \frac{\int_{m_l}^{m_u} Y_{X}\frac{dN}{dm}\,dm}{\int_{m_l}^{m_u} \frac{dN}{dm}\,dm} $}
%\vspace{-2mm}
\tablenotetext{$\tiny$a}{{The adopted effective yields for CCSNe, POPIII-HE, and POPIII-LE assume a relatively top-heavy power-law IMF with a power index $\alpha = 2.1$. The results for $\alpha = 2.35$, 3, and 1 are also given for comparison. The CCSN yields are for a metallicity of $0.02Z_\odot$. The SMS yields are for a mass of $10^5$ {\msun} and a metallicity of $0.1Z_\odot$.}}
%\vspace{-2mm}
%\tablenotetext{$\tiny$ b}{POPIII-HE stars.}
%\vspace{-2mm}
%\tablenotetext{$\tiny$ c}{POPIII-LE stars.}
%\vspace{-2mm}
%\tablenotetext{$\tiny$ d}{For a single metal rich SMS with $M=10^5M_\odot$ \citep{Nagele+2023}.}
%\vspace{-2mm}
\end{deluxetable*}

\section{Results}
\label{sec:res}

\subsection{Model dependencies on inflow rate and timescale, outflow rate, and IMF slope}
\label{sec:dependencies}

Before we proceed to apply the model to real galaxies, we first discuss how the time evolution of 
{the O/H, C/O, and N/O ratios changes as a function of the model parameters. Specifically, in Figure \ref{fig:trends_time} we show these variations as a function of the inflow rate (first column) and timescale (second column), outflow rate (third column), and IMF slope (fourth column). As an example, we adopt a two-burst SFH, with each burst lasting 10 Myr. The SFH extends for 50 Myr, and the bursts are separated by $\sim$ 30 Myr. The inflowing gas is assumed to be pre-enriched by standard CCSNe, to an absolute level $\rm Z_{inf,O}=10^{-6}$.} We consider a grid of total infall masses ranging from $\rm log(M/M_{\odot}) = 9$ to $12.5$ in 0.5 dex increments, infall timescales ranging from $\rm log(\tau/yr) = 7$ to $8.5$ in 0.5 dex increments, outflow efficiencies  $w = 0, 100, 500$,  and IMF slopes of $\alpha=2.1, 2.35, 2.5, 2.8, 3.0$ for the stellar populations within the galaxies. To isolate the effect of each parameter, we vary one at a time while keeping the others fixed. We show the results in Figure \ref{fig:trends_time}. In general the starting value of O/H, C/O and N/O (i.e. prior to the onset of the first SF episode) are those corresponding to the IMF integrated CCSN yields, which are assumed as pre-enrichment sources in this exercise.

\begin{figure*}[htbp]
\centering
     \includegraphics[width = 1\textwidth]{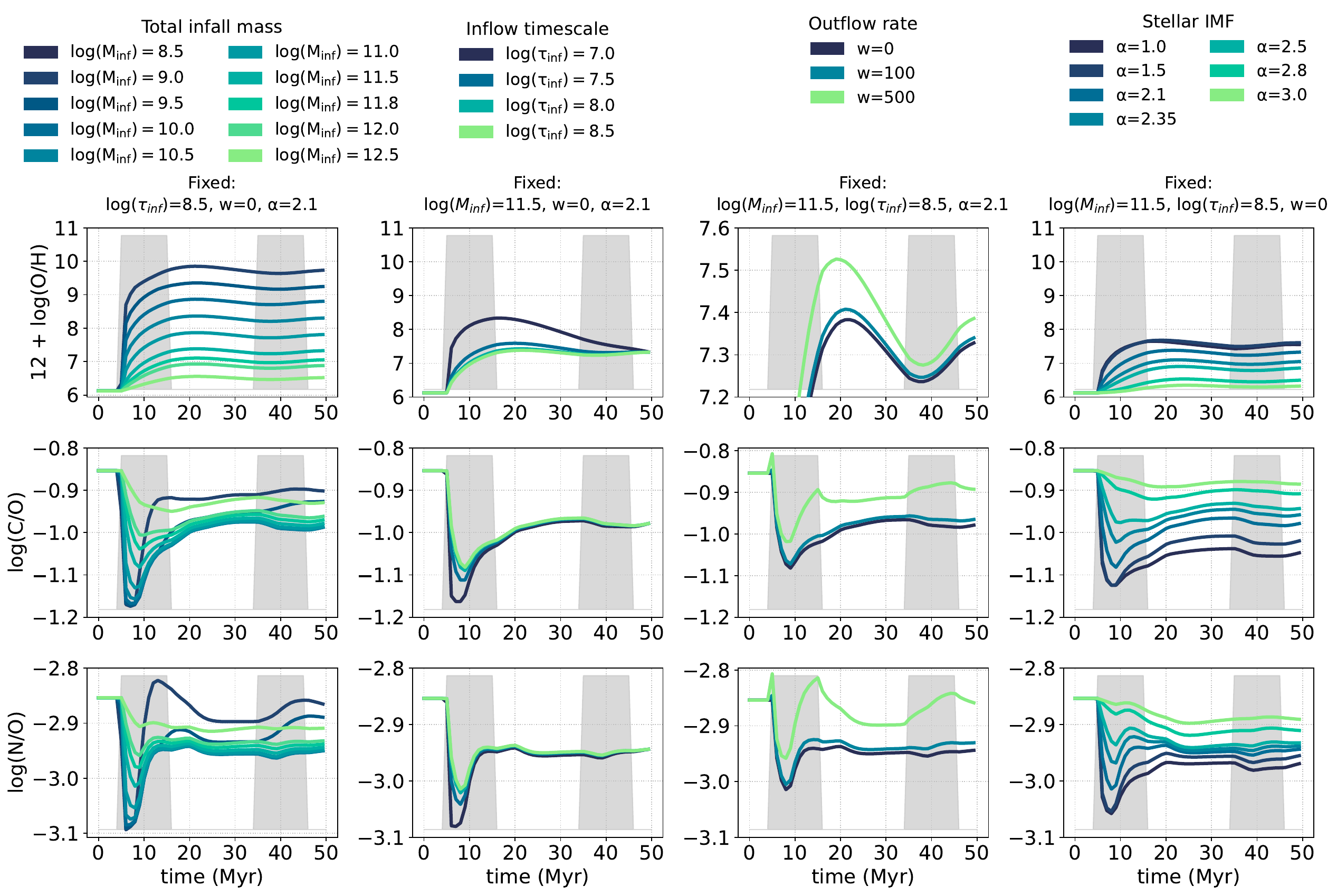}
    \caption{Time evolution of log(O/H), log(C/O), and log(N/O) as a function of the total infall mass (first column, ranging from $\rm log(M/M_{\odot}) = 9$ to $\rm log(M/M_{\odot}) = 12.5$ in 0.5 dex increments), infall timescale (second column, with values from $\rm log(\tau) = 7$ to $\rm log(\tau) = 8.5$ in 0.5 dex increments), outflow rate (third column, with values $w = 0, ~100, ~500$), and IMF slope (fourth column, with values $\alpha=2.1, ~2.35, ~2.5, ~2.8, ~3.0$). {For this exercise, we assume IGM/CGM pre-enrichment by CCSNe}. In each column we evaluate the effect of one parameter, fixing the others as described at the top of each column. The curves are color-coded from dark blue to green to indicate increasing values of each parameter. The grey shaded regions show the assumed bursty SFH, spanning 50 Myr, and characterized by two 10 Myr long bursts forming a total mass of $10^8$ $M_{\odot}$. Note the change in the y-axis range for the trend of O/H vs. outflow rate.}
\label{fig:trends_time}
\end{figure*}

The O/H abundance evolution is shown in the first row of Figure \ref{fig:trends_time}. We observe some general trends which are independent of the parameter values. Specifically, O/H has a rapid increase right after the SF onset, as massive stars form and release large amounts of oxygen into the ISM. Subsequently, O/H flattens, as the SF episode ends, stars die and the ISM is diluted by the inflow. Once the first SF episode ends, the evolution is dominated by the inflow of nearly pristine gas, which leads to a decrease in O/H, as the accreted material dilutes the existing O/H content. With the onset of the second SF episode, O/H rises again, but more subtly. This is because the gas reservoir has grown in the meantime, making the O contribution from newly formed stars less noticeable.

We observe that reducing the total infall mass (first panel) leads to higher O/H, due to reduced dilution from metal-poor infalling gas. A shorter inflow timescale results in a more pronounced increase in O/H during the first star formation episode. This is because, under our assumption of an exponentially increasing inflow rate (see Section \ref{CEM}), the dilution from accreting gas is minimal at early times. Later on, O/H drops sharply due to the larger amounts of accreted metal-poor gas. In contrast, a longer inflow timescale (second panel) leads to more gradual and steady gas accretion over time, resulting in smoother changes in oxygen abundance as the ISM is continuously replenished with low-metallicity gas.

The third panel shows that varying the outflow rate does not strongly affect O/H. A higher outflow rate only slightly increases O/H right after the first SF episode. This occurs because the oxygen produced by massive stars is released into a smaller gas reservoir, since much of the gas is removed by outflows. This effect is less noticeable during the second SF episode, when the gas mass has already increased significantly.

The fourth panel shows how \(\mathrm{O/H}\) responds to changes in the slope of the IMF of the stars formed within the galaxy. Specifically, more top-heavy IMFs lead to higher overall \(\mathrm{O/H}\) than bottom heavy IMFs, due to the larger number of massive stars formed to contribute oxygen as CCSNe.

The second and third rows of Figure~\ref{fig:trends_time} show the time evolution of \(\mathrm{C/O}\) and \(\mathrm{N/O}\). Since their evolution is similar, we describe them together in the following. Both ratios exhibit a general trend of rapidly decreasing at the onset of the first SF episode. This drop is driven by the rapid increase in \(\mathrm{O/H}\). Following this initial drop, \(\mathrm{C/O}\) and \(\mathrm{N/O}\) increase again, reflecting the flattening of \(\mathrm{O/H}\). After the first SF episodes, the ratios remain stable, as the stellar products of the first SF episode have contributed to the ISM enrichment to their full extent. As in the case of \(\mathrm{O/H}\), the impact of the second SF episode on \(\mathrm{C/O}\) and \(\mathrm{N/O}\) is milder, due to the larger gas reservoir present at that time.

We observe that increasing either the total infall mass or the inflow timescale produces a less pronounced initial drop in C/O and N/O. This is due to stronger dilution from the larger or more sustained inflow of nearly pristine gas. A similar effect is seen when we increase the IMF slope. A more bottom heavy IMF implies fewer massive stars, resulting in lower oxygen production and therefore a milder change in the abundance ratios. In contrast, we observe that increasing the outflow rate removes larger amounts of oxygen enriched gas, causing a more pronounced increase in \(\mathrm{C/O}\) and \(\mathrm{N/O}\).

{From Figure~\ref{fig:trends_time}, we see that the $\log(\mathrm{N/O})$ ratios produced by our models stay below $\sim -2.8$. These values are significantly lower than the typical $\log(\mathrm{N/O}) \sim -1.5$ observed in lower-redshift galaxies (Figure~\ref{fig:logfig}). This difference arises because our galaxies have very short SFHs, so nitrogen is supplied exclusively by massive stars. In contrast, the lower-redshift systems shown in Figure~\ref{fig:logfig} have much longer SFHs and therefore receive substantial contributions of carbon and nitrogen from AGB stars. A straightforward way to see this is that the IMF--integrated AGB yields (assuming a slope $\alpha = 2.1$) naturally produce $\log(\mathrm{N/O}) \sim -0.46$, fully consistent with the elevated $\log(\mathrm{N/O})$ ratios observed in local galaxies. In our exercise, where only CCSNe are included as enrichment sources, the predicted N/O values remain low; additional nitrogen-producing channels would increase the resulting N/O ratios. We also remind that the CCSNe yields we use do not include rotational mixing, which can contribute to increase the N/O ratio in old Milky Way halo stars \citep[e.g.,][]{Spite+2021}.} {In addition, failed CCSNe (not included here) may be another source for increasing N/O ratios at high redshift (e.g., \citealp{VincenzoKobayashi2018} and references therein).}

\begin{deluxetable*}{lcccccr}[ht]
\label{tab:results}
\tabletypesize{\footnotesize}
\tablecolumns{7} 
\tablecaption{\label{tab:res} Main results obtained from our chemical evolution model.} 
\tablehead{\colhead{Target} & \colhead{$\rm M_{\rm inf}$ [\msun]} & \colhead{$\rm Z_{inf,O}$} & \colhead{$\rm \tau_{\rm inf}$[yr]} &  \colhead{$\alpha$} & \colhead{IGM pre-polluter} & \colhead{predicted log(N/O)}}
\startdata
ERO s04590$^{a}$ & $10^{9.0}$ & $10^{-4}$ & $10^{6.5}$ &  1.0 &  CCSNe & $-2.87$ \\
CEERS-1019 & $10^{10.8}$ & $10^{-5}$ & $10^{6.8}$ &  1.0 &  SMSs  & 0.10\\
MACS1149-JD1 & $10^{9.8}$ &  $10^{-6}$ & $10^{7.5}$ & 1.0 &  CCSNe/POPIII-HE & $-2.92~(-2.79)^{b}$\\
JADES-GS-z9-0 (N/O detection) & $10^{10}$ & $10^{-5}$ &  $10^{7.5}$ &  2.1 & CCSNe & $0.17$ \\
JADES-GS-z9-0 (N/O nondetection) & $10^{10}$ & $10^{-6}$ &  $10^{8.5}$ &  2.35 & CCSNe/POPIII-HE & $-2.86$ \\
MACS0647–JD$^{a}$ & $10^{8.8}$ & $10^{-5}$ & $10^{8.2}$ &  3.0 &  SMSs & 0.15 \\
GN-z11 & $10^{9.2}$ & $5\times 10^{-4}$ & $10^{7.5}$ &  2.8 &  SMSs & 0.94 \\
GS-z12$^{a}$ & $10^{8.2}$ & $10^{-5}$ & $10^{8.5}$ & 3.0 &  SMSs & 0.61 \\
JADES-GS-z14-0$^{a}$ & $10^{9.5}$ & $10^{-5}$ & $10^{6.8}$ & 1.0 &  CCSNe & $-2.70$\\
\enddata
%\vspace{2mm}
\tablenotetext{$\tiny$ a}{Best fit obtained considering only O/H and C/O data.}
\vspace{-2mm}
\tablenotetext{$\tiny$ b}{The two values correspond to the best-fit results obtained when considering CCSNe (POPIII-HE) as the primary sources of IGM/CGM enrichment.}
\vspace{-2mm}
\end{deluxetable*}

\subsection{Application to the sample galaxies}
\label{sec:CO_NO}

In this Section we show the results obtained by applying our chemical evolution model to individual galaxies. The best fit model is determined through a $\chi^2$ minimization considering all the observational constraints:  O/H, C/O, and N/O (or O/H and C/O, when N/O is not available). Abundance measurements available only as upper or lower limits are excluded from the $\chi^2$ calculation, but are still used to constrain the best-fit solution. For each galaxy, we run the chemical evolution model assuming its specific SFH, and compare the results from the final step of the calculation  (i.e., at the galaxy’s age) with the observed values. 
We  verify that the total baryonic mass predicted by our models, i.e. $M_s + M_g$ in our equations, remains within the uncertainties of the estimated baryonic mass of the galaxy, defined as $M_{bar} = \frac{f_b}{\rm STHM} \times M_\star$, where $f_b=0.16$ is the cosmic baryon fraction \citep{Plank2020} and $\rm STHM=0.005$ is the average stellar-to-halo mass ratio at \( z \sim 8 \) for galaxies of stellar masses similar to ours \citep{Behroozi+2013}.

In the following, we present our results for CEERS-1019 in details, while the results for the other galaxies are shown in Table~\ref{tab:results} and Appendix \ref{sec:appendix_model}. For CEERS-1019,  measurements  of both C/O and N/O are available and  shown as white stars with corresponding uncertainties in Figure~\ref{fig:CO_NO_CEERS}. This figure also shows the C/O and N/O vs. 12 + log(O/H) computed for different sources of IGM/CGM pre-enrichment, color-coded as follows: {dark blue for SMS, light blue for POPIII-LE, violet for POPIII-HE, and light green for CCSNe}. The curves show the full time evolution of each calculation, with the endpoint (corresponding to the age of the galaxy, $t_{obs}$) represented by filled circles. We mark the best fit model with thicker curves.

{We observe that our best‑fit values for C/O and N/O at $t_{obs}$
are consistent with the measurements within the associated uncertainties.} The best fit model for this galaxy predicts a total infall mass $M_{\inf} = 10^{10.8}$ \msun, an infall timescale of $\tau_{\rm inf}= 10^{6.8}$ yr, and IMF slope of the  stellar populations $\alpha = 1.0$. The chemical abundances of this galaxy can only be reproduced by an IGM/CGM pre-enriched by SMSs and with an absolute inflow metallicity $\rm Z_{inf,O} = 10^{-5}$. All other models severely under-predict the observed N/O abundance while at the same time over-predicting the observed 12+log(O/H). 

CEERS-1019 is not unique  in the need for SMSs (see Appendix \ref{sec:appendix_model}). We find that MACS0647--JD, GN-z11, and GS-z12 also require an inflow of gas pre-enriched by SMSs (with an absolute oxygen enrichment level exceeding \(10^{-5}\)) in order to reproduce their observed chemical abundances. In contrast, the abundance patterns observed in ERO~s04590, MACS1149--JD1, and JADES-GS-z14-0 are consistent with inflows pre-enriched by standard CCSNe.

{From the results shown in this Section and in Appendix \ref{sec:appendix_model}, we observe that the best-fit $Z_{\rm inf}$ for our galaxies are consistent with the $Z_{\rm IGM}$ found at $z\sim8-9$ \citep[e.g.,][]{DeCia+2018, Tanvir+2021}.}

{Since most galaxies in our sample lack a $\log(\mathrm{N/O})$ measurement (see Table \ref{tab:highzCO} and Section \ref{sec:appendix_model}), we tested whether the best--fit model obtained for CEERS--1019 is recovered when only the $\log(\mathrm{C/O})$ ratio is used as a constraint. In this restricted case, the inferred best-fit parameters shift to 
$M_{\mathrm{inf}} = 10^{10}\,\mathrm{M_\odot}$, 
$\tau_{\mathrm{inf}} = 10^{7.5}\,\mathrm{yr}$, 
$\alpha = 2.5$, 
and $Z_{\rm {inf,O}} = 10^{-4}$. 
Moreover, the preferred pre-enrichment source is no longer SMSs, but Pop~III-LE stars. This demonstrates that simultaneous C/O and N/O measurements are essential for reliably identifying a galaxy's chemical evolution history.
}

 %{plot_main_MC24_new_oct25.pdf}
 %{plot_CO_NO_ceers1019_inset.pdf}
\leavevmode
\begin{figure*}[htbp]
\centering
     \includegraphics[width = \textwidth,  trim = {0em, 6em, 0em, 0em}]{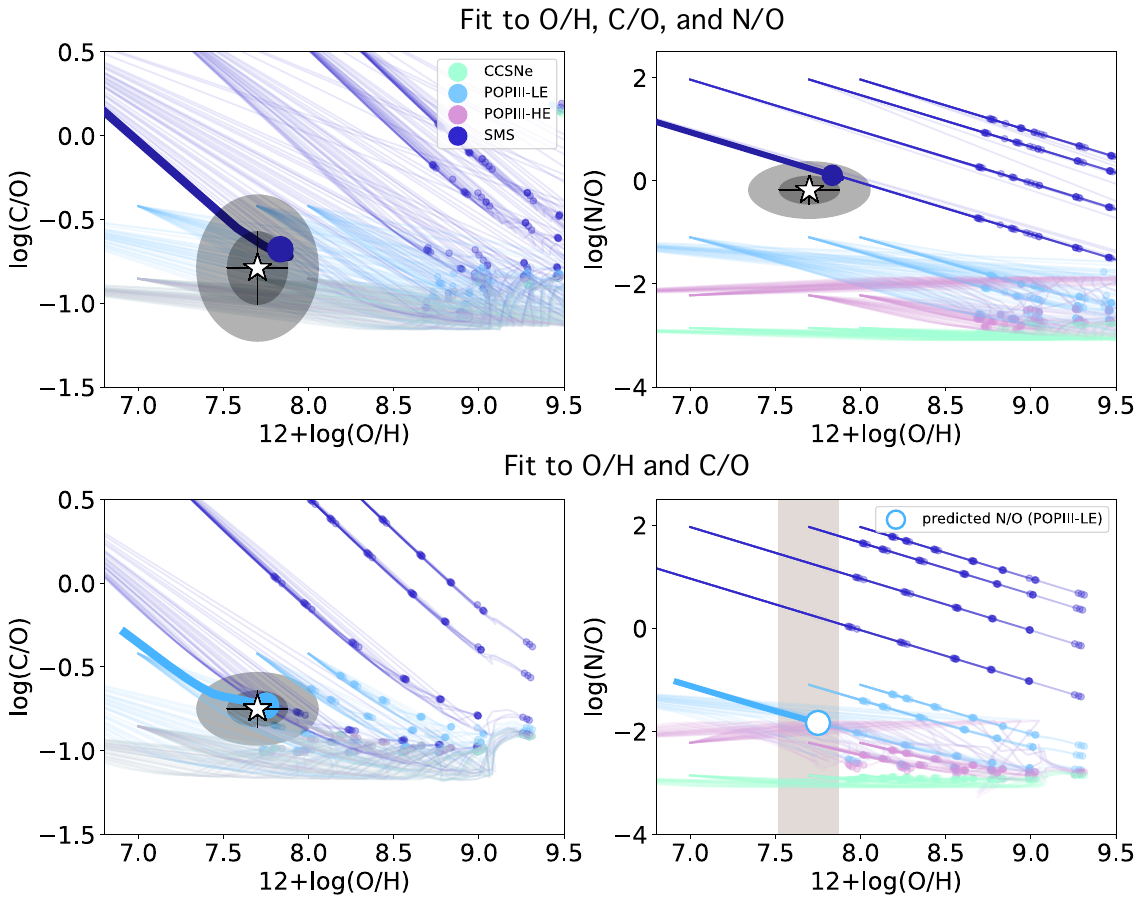}
    \caption{Observed and predicted O/H, C/O, and N/O values for CEERS-1019. The observed values are marked with star symbols together with the corresponding uncertainties. The colored curves  are predicted by our models. Different color symbols represent different IGM/CGM pre-enrichment: SMSs (dark blue), CCSNE (light green), POPIII-LE stars (light blue), and POPIII-HE (violet). The values at $t_{obs}$ are marked with small circles. Best fit models are shown with thicker curves and bigger $t_{obs}$ circles. {The top panels compare the models with measurements of O/H, C/O, N/O. The gray ellipses indicate the $1\sigma$ and $2\sigma$ error boundaries.} We observe that the best fit model associated with IGM/CGM pre-enrichment by SMSs (thick blue curve) is the only one consistent (at $t_{obs}$) with the observed values within their uncertainties. The bottom panels show the results - and best fit model - obtained under the assumption that CEERS‑1019 is constrained only by its O/H and C/O measurements.}
\label{fig:CO_NO_CEERS}
\end{figure*}

\vspace{3mm}

\subsection{Star formation efficiency}
\label{sfe}

In this Section we use the best-fit parameters to compute a gas depletion {rate},  which we define using the best-fit gas mass and the measured SFR at the age of the galaxy ($t_{\rm obs}$): $\rm \Gamma_{dep,g}  =  SFR_{obs}/M_{g}^{best-fit}$. This is defined as {the inverse} of the time it would require to consume all the gas if the SFR were to remain constant at the observed value. We also compute an integrated star formation efficiency, defined as $\rm SFE_{\rm integrated} = M_{\star,obs}/(M_{g}^{best-fit} + M_{\star,obs})$, where $\rm M_{\star,obs}$ is the observed stellar mass (see Table \ref{tab:highzCO}). SFE$_{\rm integrated}$ represents the average gas-to-star conversion efficiency over the (short) lifetime of each galaxy. Figure \ref{fig:sfe} shows how the depletion {rates} and integrated star formation efficiencies at $z\gtrsim 8$ depend on stellar mass. We also present the comparison with both the Milky Way and a sample of $0\lesssim z \lesssim 1$ galaxies.  

{Among the relatively small sample of high-redshift galaxies studied, there is no clear trend with redshift for either the depletion rate or the integrated SFE. Identification of any such trend requires data on more such galaxies. On the other hand, compared to low-redshift galaxies, high-redshift ones tend to occupy different regions on the depletion rate vs. stellar mass and the integrated SFE vs. stellar mass planes. Specifically, for the same stellar mass range, high-redshift galaxies typically have much higher depletion rates (shorter depletion times). In addition, they tend to have much lower integrated SFEs and smaller stellar masses. Both features might simply reflect that they are gas-rich systems actively forming stars, in contrast to low-redshift systems where star formation is mostly completed.}

% At $z>8$ the depletion time scale is on average $\approx 100$ times shorter than that of the Milky Way and of local galaxies, and more consistent with the timescales observed at $z\sim1$. 
The integrated SFEs of high-redshift galaxies are on average a factor of 10 lower than local galaxies of similar masses, and a factor of $\approx 5$ lower than  galaxies at $z\lesssim1$. %This suggests that galaxies at $z>8$ are converting gas into stars more efficiently than lower redshift galaxies, \textcolor{mycolor}{continuing the trend observed at $z\sim0$ and $z\sim1-2$}. However, being very young, did not have time to increase their integrated SFE. 
As the color of the symbols for $z\gtrsim 8$ galaxies show, we observe no redshift dependence among our galaxy sample, suggesting a consistent mode of star formation from $z\sim8$ to $z\sim14$. The small number of galaxies, however, may hide redshift trend.

\begin{figure*}[htbp]
\centering
\includegraphics[width =1 \textwidth]{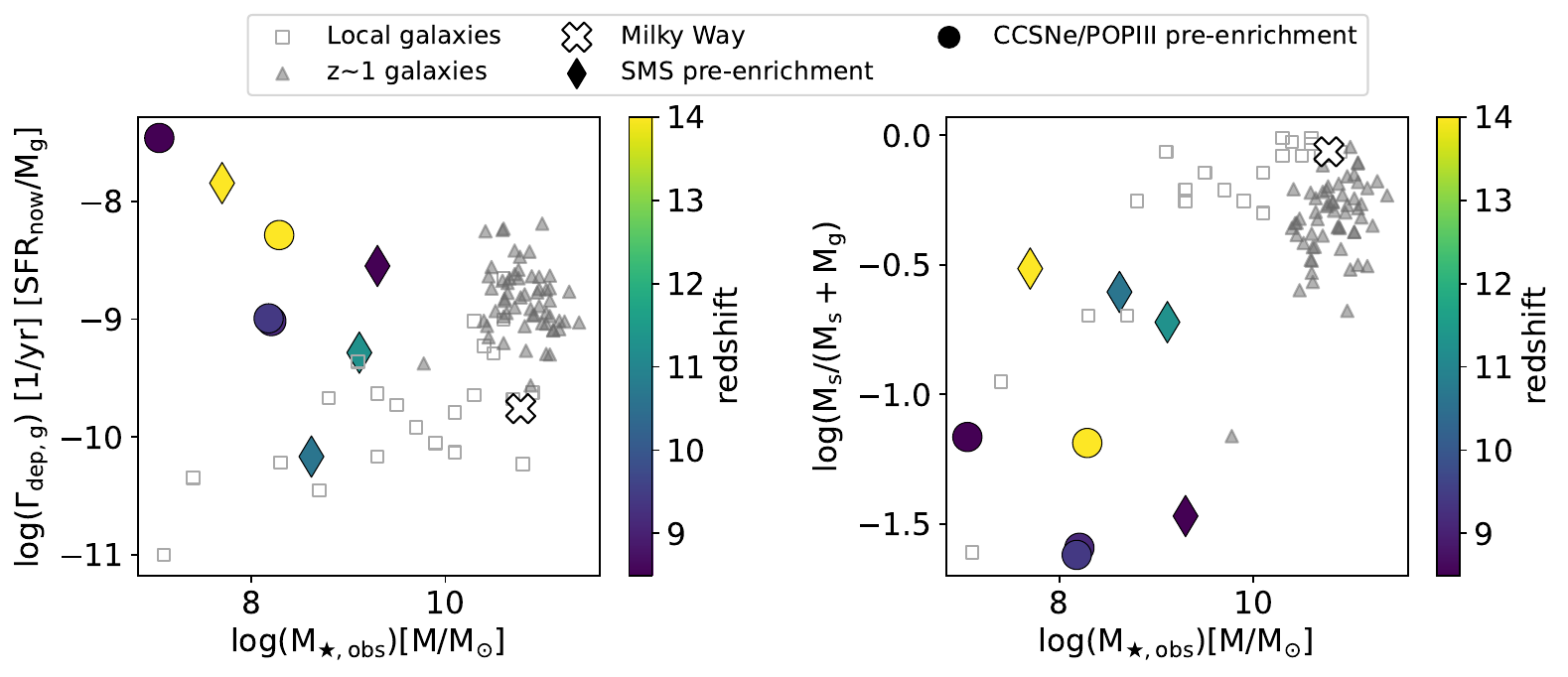}
\caption{Gas depletion {rates} and integrated star formation efficiencies as functions of stellar mass and redshifts for the galaxies analyzed in this work compared to the Milky Way ($X$-shaped marker) and samples of local galaxies \citep{Leroy+2008} (small squares) and $z\sim1$ galaxies \citep{Tacconi+2013} (small triangles). {Colored circles and diamonds differentiate between galaxies with inflow pre-enrichment by CCSNe/POPIII and SMS stars, respectively.} %{The arrow (on the left) and pentagon (on the right) mark the location of Pox 186, assuming the values reported in \citet{Eggen+2021}. Circles and diamonds mark pre-enrichment of the IGM by CCSNe/POPIII and SMS stars, respectively, color-coded by redshift.}
}
\label{fig:sfe}
\end{figure*}

\section{Discussion}
\label{sec:discussion}

\subsection{Comparisons with the literature}

JWST allows us to derive chemical abundances in galaxies up to $z\sim14$ \citep{Carniani+2024}, and shows that early galaxies are characterized by C/O abundance ratios that are generally consistent with those of lower redshift sources with similar oxygen abundance ({\citealp[e.g.,][]{Berg+2019b} for the local Universe; 
\citealp{Deugenio+2024, Hsiao+2024, Stiavelli+2023} for high redshift}), while the N/O ratios are on average one dex (or more) higher than those observed in the local Universe {\citep[e.g.,][]{MarquesChaves2024, Curti+2025}. Although the prevalence of N‑emitters at high redshift remains to be established, it is interesting to understand which evolutionary pathways can reproduce such extreme abundance ratios.}

{The relative C/O abundance is generally linked to a galaxy's evolutionary stage or the presence of differential outflows. {High (low) C/O abundance ratios} can result from older (younger) stellar populations, depending on whether AGB stars have had sufficient time to contribute carbon to the ISM. Additionally, the oxygen content of galactic outflows expelled into the IGM/CGM can lead to a broad range of C/O ratios \citep{Berg+2019b, Rizzuti+2024}.}
{More exotic stellar populations, such as POPIII stars or WR stars, as well as top-heavy IMFs and complex SFHs, have also been proposed to explain the highest observed C/O abundance ratios. WR stars, for example, have been suggested as a possible explanation for GN-z11’s location in diagnostic diagrams involving carbon emission lines \citep{Gunawardhana+2025}. A top-heavy IMF, combined with WR stars or pair-instability supernovae (a class of POPIII stars), can instead reproduce the low C/O abundance ratio observed in JADES-GS-z9-0 (under a single-burst SFH scenario \citealp{Curti+2025}). In the case of GS-z12, Pop III-LE stars have been proposed as a potential source of the high observed C/O ratio \citep{Deugenio+2024}.}
%O-rich outflows have also been proposed to explain elevated C/O at low O/H \citep{Berg+2019b, Rizzuti+2024}. %, but our results, along with those of \citet{Bassini+2023}, indicate that low outflow rates are required to match the observed O abundances, particularly at high $z$.

Four out of the eight analyzed galaxies have measured N/O abundances. CEERS-1019 and GN-z11 show elevated values ($\log(\text{N/O}) > -0.36$), MACS1149--JD1 has $\log(\text{N/O}) \lesssim 0$, and JADES--GS--z9--0 exhibits a $\sim$ solar-like N/O. Several mechanisms have been proposed to explain high N/O at high redshift {at a fixed metallicity}, including pristine gas inflows, WR star enrichment, and O-rich outflows \citep{Stiavelli+2023}. \citet{Rizzuti+2024} found that a total infall mass of $10^9\,M_\odot$, high SFE, and O-rich outflows can reproduce the observed N/O in galaxies at $z\sim3$--14. Other scenarios involve POPIII stars, tidal disruption events, ejecta from very massive stars formed via stellar collisions, and SMSs \citep{Cameron+2023b, Charbonnel+2023, Nagele+2023b, Watanabe+2024}. Top heavy IMFs \citep[e.g.,][]{Curti+2025} and bottom heavy IMFs \citep{ArellanoCordova+2024} have also been considered. \citet{KobayashiFerrara2024} suggest that the supersolar N/O in GN-z11 can be explained by a finely tuned SFH with intermittent bursts and WR nucleosynthesis following the second burst. SMSs formed within the galaxy have also been proposed as contributors to N enhancement in GN-z11 \citep{Charbonnel+2023, Nagele+2023a}. \citet{Maiolino+2024} argue that N enrichment in GN-z11 is likely confined to its nuclear region near the central black hole, and that He~II emission in its halo may indicate the presence of POPIII stars \citep[e.g.,][]{Maiolino+2024, Maiolino+2026}.

{In this work, we consider another factor that must be accounted for when studying chemical enrichment at early times: the enrichment level and chemical composition of the gas flowing into galaxies, which triggers star formation. {This aspect becomes particularly important at early times, when the short SFHs of young galaxies are not sufficient to erase the chemical imprint of the pre‑enriched inflowing gas, allowing its abundance pattern to remain visible in the ISM.} %\textcolor{blue}{{This needs to be changed}: The first aspect making the chemical enrichment of inflows important is the fact that, as shown in \citet{Bassini+2023}, inflow rates dominate over outflow rates during early epochs.}

We find that, when the enrichment level and chemical composition of the inflowing gas are properly considered, 
%they play a larger role than the IMF slope and outflow rate in reproducing the observed chemical abundances in the analyzed galaxies. Accounting for the CGM pre-enrichment, we find that 
the observed C/O and N/O abundance ratios of half of the analyzed sample of $z>8$ galaxies can only be simultaneously reproduced if the inflowing gas has been previously enriched by SMSs, with an absolute oxygen enrichment level of $\rm Z_{inf,O} \gtrsim 10^{-5}$.} The relative abundances of the other half of the sample can be reproduced with pre-enrichment by either CCSNe or POPIII stars. 

{Interestingly, for ERO~s04590, which is the lowest-redshift galaxy in our sample, our model predicts a higher absolute enrichment level of the inflowing gas of $Z_{\rm inf,O} =10^{-4}$, possibly due to the natural increase in IGM enrichment towards lower redshifts. Among the galaxies, only MACS1149-JD1 appears to have formed from a more primordial gas or a metal poor gas pocket with $Z_{\rm inf,O} \approx 10^{-6}$, enriched by either CCSNe or POPIII-HE \citep[e.g.,][]{Tornatore+2007, Pallottini+2014a, Venditti+2023, Venditti+2024a, Katz+2023}.
}

\subsection{On the contribution of metal enriched Supermassive Stars}
\label{sec:sms}

Literature studies, such as those by \citet{Nagele+2023a} on GN-z11, suggest that metal enriched SMSs with $Z\sim0.1~Z_{\odot}$ are responsible for the high observed N/O abundances.
However, while GN-z11 reaches the $\gtrsim0.1~Z_{\odot}$ neeeded to trigger the formation of metal enriched SMS through gas-rich, metal-rich galaxy mergers \citep[e.g.,][]{Mayer+2010, MayerBonoli2019, Nagele+2023b}, most young N-loud galaxies (e.g., CEERS-1019, \citealp{MarquesChaves2024}) never reach  $Z\sim0.1~Z_{\odot}$. {According to our model, where the pre-enrichment of inflowing gas plays a key role}, the metal rich SMSs form in dense environments, where 
stellar collisions or gas-rich mergers facilitate their formation, and subsequently explode, enriching the surrounding IGM/CGM. This pre-enriched gas is later accreted by the galaxies, triggering star formation. Given the short SFHs that characterize early galaxies, the chemical signature of the CGM/IGM remains imprinted in their observed ISM abundances.

Among the galaxies in our sample, we find that CEERS-1019 and GS-z12 do not reach a high enough gas metallicity to be able to form SMSs in situ. Therefore, they require pre-enrichment of IGM/CGM by SMSs in order to reproduce their observed abundances. % In contrast, ERO~s04590 and JADES-GS-z9-0 do not need pre-enrichment by SMSs, but instead CCSNe or POPIII stars are sufficient. These scenarios can be confirmed with measurements of N/O abundances.

\subsection{High SFE at the Cosmic Dawn}

JWST has uncovered a growing population of massive galaxies at $z>8$ (i.e., the Cosmic Dawn), many with $M_{\star}> 10^8$ \msun\ and SFR $\gtrsim$ 1 \msun $\rm~yr^{-1}$ \citep[e.g.,][]{Finkelstein+2023, Labbe+2023, Carniani+2024}. These galaxies appear more abundant than predicted by pre-JWST galaxy formation models, particularly at the bright end of the UV luminosity function \citep{Boylan-Kolchin2023}. Several scenarios have been proposed to explain the presence of these galaxies in the early universe: feedback-suppressed models \citep{Dekel+2023, Andalman+2025}, systematic biases in UV luminosity estimates (possibly due to low mass-to-light ratios), AGN contamination, top-heavy IMFs, unusual dust properties \citep{Mason+2023,Cameron+2024}, bursty SFHs (where recent starbursts temporarily boost UV brightness-- \citealp{Shen+2023}), attenuation-free models \citep[e.g.,][]{Ferrara+2025, Ferrara+2026}, or non-standard cosmologies (such as Early Dark Energy or primordial black hole seeds, which could increase the number of massive halos--\citealp{Klypin+2021, LiuBromm2022}). 

According to the high SFE hypothesis, galaxies at $z\gtrsim8$ are characterized by elevated baryonic surface densities, which suppress the stellar feedback mechanisms usually responsible for regulating star formation. This suppression allows gas to convert into stars more efficiently, resulting in significantly enhanced SFE during the early stages of galaxy formation \citep{Dekel+2023, Andalman+2025}. 

In support of this scenario, our analysis in Section~\ref{sfe} suggests that the $z\gtrsim8$ galaxies are converting gas into stars very efficiently {(i.e., with high gas depletion rates) close to the time of observation}. We also find that the average past efficiency of converting baryons into stars (integrated  SFE in Figure~\ref{fig:sfe}) is lower than that of $z\sim 1$ massive galaxies, {which may simply reflect that these high-redshift galaxies have acquired a lot of gas recently to boost their star formation}.  We also find that to interpret the chemical abundances of the majority of these systems we need low/negligible outflow rates.
Taken together, these results paint a picture in which the $z\gtrsim 8$ galaxies in our sample are observed in a phase of enhanced SF activity without strong feedback, while still having retained a large reservoir of gas, possibly as a result of the large inflow rates.

\subsection{Challenges in modeling and observing chemical enrichment in early galaxies}
\label{sec:need}

One of the key insights from our study is that the C/O vs. 12 + log(O/H) plane alone is insufficient to determine the most plausible enrichment scenario for early galaxies (see Section \ref{sec:CO_NO} and Figure \ref{fig:CO_NO_CEERS}). In contrast, the addition of the N/O vs. 12 + log(O/H) plane provides much more stringent constraints. Thus, both C/O and N/O ratios are  necessary to accurately identify early-time enrichment. {In a future work (Citro et al., in prep.), we will investigate whether incorporating additional relative abundance ratios can provide more precise constraints on early‑time enrichment pathways and star formation histories. We will also expand the set of stellar sources formed after the inflow‑triggered star formation. Including ratios such as Ne/O offers a way to quantify the relative contributions of different massive‑star channels. For example, SMS enrichment is expected to produce high N/O and high Ne/O ratios because O is strongly depleted, whereas enrichment dominated by very massive stars (\(M>30\,M_\odot\)) produces both low Ne/O \citep{Isobe+2023, Watanabe+2024} and low N/O. Applying our chemical evolution model to both observed abundance ratios simultaneously using the observed SFH of the galaxy will help identify the dominant enrichment channel and also assess whether a given high-redshift system could be a progenitor of present-day globular clusters (expected to have high N/O, \citealp[e.g.,][]{Senchyna+2024}) with higher accuracy. Moreover, comparing our models to both abundance ratios provides a way to test more complex star formation histories that may not be captured by photometry-derived SFHs. This is particularly relevant for systems showing high N/O but low Ne/O, where a bursty or multi-episode SFH with multiple enrichment channels may be required.}

%Furthermore, comparing our models to multiple abundance ratios provides a means to test more complex star formation histories that may not be captured by photometry‑derived SFHs. This is particularly relevant for systems exhibiting high N/O but low Ne/O, where a bursty or multi‑episode SFH with multiple enrichment channels may be required.

Beyond these future extensions, several assumptions in our current modeling framework warrant further consideration. {First, we are assuming the ISM is enriched through a typical gas-star-gas cycle. However, only CEERS-1019, MACS1149-JD1, and JADES-GS-z14-0 show a clearer star-forming nature, while the other targets in the sample might host an AGN. {The AGN presence could influence how heavy elements are retained in the ISM, thereby affecting the N/O ratio. For example, \citet{Isobe+2025} reported elevated $\log(\mathrm{N/O}) > -0.6$ in broad-line AGN stacks (assuming $n_e\sim300\,\rm cm^{-3}$).} %However, \citet{Hayes+2025} ruled out AGN activity as the source of N/O enhancement by analyzing galaxy stacks at $z \sim 4\text{--}7$.}
%They argue that the increased N/O in AGNs might be due to the presence of dense nuclear star formation, which traps nitrogen-rich gas, leading to N over-enrichment.}

Second, although our models rely on the most realistic SFH available for the considered targets, these are photometry-derived SFHs. Accurately determining galaxy SFHs from photometry is challenging due to the age-metallicity-dust degeneracy in galaxy spectra \citep[e.g.,][]{Conroy2013} and the complex nature of spectral fitting, which often depends heavily on prior assumptions \citep[e.g.,][]{Ocvirk+2006, Leja+2019}. These difficulties suggest that the stellar ages for $z\sim10$ galaxies, as derived from current data, remain highly uncertain \citep{Tacchella+2022}.

{Another notable aspect is that the high nitrogen abundances at high redshift are most often inferred from UV nitrogen lines such as \ion{N}{3}]~$\lambda\lambda1747,1749$, \ion{N}{4}]~$\lambda1486$, or \ion{N}{4}]~$\lambda1483$ \citep{Bunker+2023, Isobe+2023, MarquesChaves2024, Schaerer+2024, Topping+2024a, Topping+2024b, Curti+2025}. However, optical nitrogen diagnostics do not always agree with the UV-derived high N/O. For instance, \citet{Stiavelli+2025} analyzed the N abundance of galaxies at $3 < z < 6$ using the optical [\ion{N}{2}]~$\lambda\lambda6548,6583$ lines and found systematically lower N/O ratios than those inferred from UV nitrogen lines for galaxies at similar redshifts. On the other hand, \citet{Berg+2025} recently reported very high N/O ratios from the optical [\ion{N}{2}]~$\lambda6584$ line for a galaxy at $z = 6.1$, comparable to the values obtained from UV nitrogen diagnostics.}

{Furthermore, the derived chemical abundances are highly sensitive to the assumed electron density. In particular, adopting a low gas density can lead to underestimations of gas-phase metallicities and overestimations of N/O ratios \citep[e.g.,][]{Hayes+2025, Martinez+2025}.}

\section{Conclusions}
\label{sec:conclusions}

In this paper, we have developed a new chemical evolution model to explore the contribution of pre-enriched inflow to the relative abundances of C, N, and O in early galaxies, and their evolution with time. Our work is motivated by the following considerations: {(1) inflow rates dominate over outflow rates when halos hosting the galaxies have acquired sufficiently deep potential wells \citep[e.g.,][]{Bassini+2023}}, and (2) given the short timescales involved at high redshift, the chemical signatures of inflowing gas can remain imprinted in the ISM abundances despite subsequent star formation. As sources of IGM/CGM pre-enrichment we consider four options: CCSNe, POPIII-HE, POPIII-LE, and SMSs.

We use our model to simultaneously reproduce the observed O/H, C/O and N/O values in a sample of eight galaxies at $z\gtrsim 8$ for which the SFH is known and used in the calculation of the chemical evolution.  
Our main findings can be summarized as follows:

\begin{itemize}

\item[-] We find that observations of  log(C/O) and  12 + log(O/H)  alone are insufficient to distinguish between different chemical evolution scenarios, and adding log(N/O) provides constraints on the enrichment sources;

\item[-] We find that no single enrichment mechanism can account for the abundance patterns of all galaxies in our sample. Some systems require accretion of gas pre-enriched by SMSs in order to reproduce their observed abundances, whereas others are consistent with enrichment from either CCSNe or Pop~III stars. {It is important to note, however, that most of the analyzed galaxies have only upper or lower limits on $\log(\mathrm{N/O})$ (and in some cases on $\log(\mathrm{C/O})$), and our models are consistent with these limits. For such systems, robust detections of $\log(\mathrm{N/O})$ are essential for drawing firm conclusions about their enrichment histories.}

\item[-] For most galaxies, we find that the infalling material was enriched to an oxygen  mass fraction of \(Z_{\text{inf,O}} \sim 10^{-5}\). Notably, ERO s04590 requires  a more metal-rich IGM/CGM, with \(Z_{\text{inf,O}} \sim 10^{-4}\), which may be linked to its lowest redshift, compared to the other galaxies. In contrast, MACS1149-JD1 originated from a significantly more metal-poor environment, with \(Z_{\text{inf,O}} \sim 10^{-6}\), consistent with the presence of POPIII stars. These values represent the enrichment level of the accreted material coming from the CGM/IGM. Observations of high redshift absorbers find $\rm -4 \lesssim log(Z_{IGM})\lesssim-3$ at $z\approx 5$, and extrapolated values of $\rm -5 \lesssim log(Z_{IGM})\lesssim-4$ at $z\sim8-9$ \citep[e.g.,][]{DeCia+2018,Tanvir+2021}, consistent with the values we find. 

\item[-] The $z>8$ galaxies in our sample appear to be undergoing intense star formation while retaining substantial gas reservoirs, likely sustained by high inflow rates in early-forming, overdense regions. Their observed chemical abundances further suggest low or negligible outflow rates, consistent with a feedback-suppressed regime under high SFE with high gas depletion rates.

\item[-] {Our results suggest that the chemical enrichment of the high‑redshift IGM and CGM is strongly shaped by local sources. At early times, evolutionary and dynamical timescales are extremely short, so there is no single dominant enrichment channel. Instead, the early phases of galaxy evolution are governed by rapid, localized processes that collectively shape the observed abundance patterns.}

\end{itemize}

\begin{acknowledgments}
A.C. thanks K. B. Mantha, Z. Martinez, D. Nandal, and M. Stiavelli for insightful conversations and discussions. The authors acknowledge the use of Copilot GPT‑5.6 for language editing. This research is based on observations made with the NASA/ESA/CSA James Webb Space Telescope obtained from the Space Telescope Science Institute, which is operated by the Association of Universities for Research in Astronomy, Inc., under NASA contract NAS 5–26555. These observations are associated with program(s) GO-6073.
\end{acknowledgments}

\vspace{5mm}
\facilities{JWST/NIRSpec, JWST/PRISM}


\software{Astropy \citep{Astropy2013}}

\appendix
\section{Model application to additional galaxies}
\label{sec:appendix_model}

Table \ref{tab:res} summarizes the results obtained for all the galaxies studied.
In this Appendix we show in more detail the results of our chemical evolution model when applied to ERO s04590, MACS1149-JD1, JADES-GS-z9-0, MACS0647–JD,  GN-z11, GS-z12, and JADES-GS-z14-0. Our results are shown in Figures \ref{fig:main_res_1}, \ref{fig:curti24_2}, and \ref{fig:main_res_2}. In these plots, nondetections are shown as gray shaded upward/downward arrows. For galaxies without an N/O estimate, we still show the range within which a potential N/O measurement would fall as yellow shaded rectangles. It is interesting to observe that models assuming different IGM/CGM pre-enrichment sources are much better separated in the log(N/O) vs. 12 + log(O/H) plane than in the log(C/O) vs. 12 + log(O/H) plane (as discussed in Section \ref{sec:need}).

For ERO s04590, the model that best reproduces the observed C/O ratio features a total infall mass of $10^{8.2}$ \msun, a very short infall timescale of \( 10^6 \) years, weak outflows, and a stellar IMF with slope \( \alpha = 2.35 \). Additionally, an IGM/CGM enriched primarily by CCSNe and characterized by an absolute oxygen metallicity of \( Z_{\text{inf,O}} = 10^{-4} \) is required.

The observed abundances of MACS1149-JD1 can be reproduced by a model assuming $M_{\text{inf}} = 10^{9.8}$ \msun, weak outflows, $\tau_{\text{inf}} = 10^{7.5}$ yr, and a top heavy IMF with $\alpha = 1.0$ for the formed stars. To match both the detected C/O ratio and the upper limit on N/O, the model requires an IGM/CGM with \(Z_{\text{inf,O}} = 10^{-6} \), enriched primarily by either CCSNe or Population III stars with high-energy explosions.

JADES-GSz9-0 represents an exception. For this galaxy, \citet{Curti+2025} report two N/O estimates: 
a detection {with fiducial value} from \ion{N}{3}]~$\lambda1750$ {(PRISM measurement)} 
and a $3\sigma$ upper limit based on the {\ion{N}{3}]~$\lambda$1747--1754 multiplet (see Table~\ref{tab:highzCO}) from the G140M grating}.

Reproducing the upper limit requires IGM pre-enrichment by CCSNe or POPIII-HE stars (see Figure \ref{fig:main_res_2}), assuming \(Z_{\text{inf,O}} = 10^{-5}\,Z_{\odot}\) and a standard IMF with \(\alpha = 2.35\). {However, applying a \(\chi^2\) fitting approach to the N/O detection using the same model parameters produces a best-fit solution that overestimates O/H and C/O and underestimates N/O (see Figure~\ref{fig:curti24_2}).

The observed O/H and C/O abundances of MACS0647-JD can be reproduced by an outflow rate consistent with $\sim0$, $\rm M_{\rm inf}=10^{9.2}$ \msun, $\rm \tau_{\rm inf}=10^{8.8}$ yr, and SMS pre-enrichment. This best fit model predicts a high log(N/O) $=0.27$. Note that \citet{Hsiao+2024} identified a young mass-weighted age of $\sim$ 50 Myr for MACS0647-JD and found its high C/O abundance surprising. However, the young mass-weighted age of the galaxy is due to the majority of its mass being produced within the last $\sim 50$ Myr of its SFH (Figure \ref{fig:sfh}). As illustrated in Figures \ref{fig:sfh}, this galaxy's more extended SFH actually allows intermediate-mass stars to leave the main sequence and contribute to the observed C/O abundance. Despite having AGB contributions to C, the presence of intermediate-mass stars alone is not sufficient to account for the observed C/O ratio, unless pre-enrichment by SMSs is considerd (see Figures \ref{fig:main_res_1} and \ref{fig:main_res_2}).

The lower limits on the C/O and N/O abundances derived for GN-z11 \citep{Bunker+2023} are reproduced by a model with an outflow rate consistent with zero, a total infall mass $\rm M_{\rm inf}=10^{8.2}$, a $\rm \tau_{\rm inf}=10^{8.5}$ year, an $\alpha = 3.0$, and an SMS-driven IGM/CGM pre-enrichment.

The chemical abundances of GS-z12 are reproduced by models with a total inflow mass of $\rm M_{\rm inf}=10^{8.2} M_{\odot}$ with $\rm Z_{inf,O}=10^{-5}$ enriched by SMS. Concerning the ISM properties, a bottom heavy IMF with $\alpha =3.0$ is needed to reproduce the results. 

The chemical abundances in JADES-GS-z14-0 can be reproduced by a zero ($w=0$) outflow rate, $\rm M_{\rm inf}=10^{9.5} M_{\odot}$ \msun, $\rm \tau_{\rm inf}=10^{6.8}$ yr, and a top heavy IMF with $\alpha = 1.0$. This system does not have N/O measurements, but its C/O ratio is most compatible with an IGM/CGM pre-enriched by CCSNe. 

{All the galaxies shown in this Appendix have only upper or lower limits on log(N/O) (and in some cases log(C/O)), and our models are consistent with those limits. For these systems, more robust detections of log(N/O) are essential to further test our models and to draw firm conclusions on their enrichment scenarios. }

\begin{figure*}[ht]
    \includegraphics[width = 0.9\textwidth]%{plot_main_AC22_new_oct25.pdf}
    {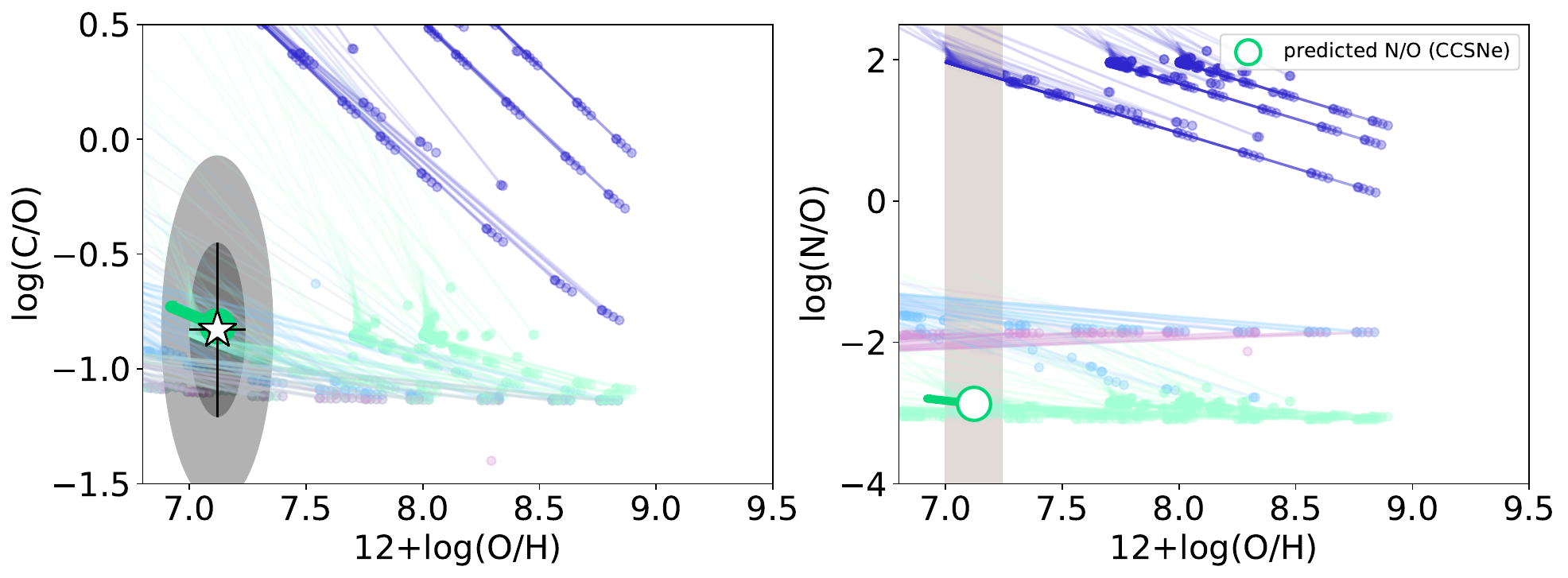}
    \includegraphics[width = 0.9\textwidth]{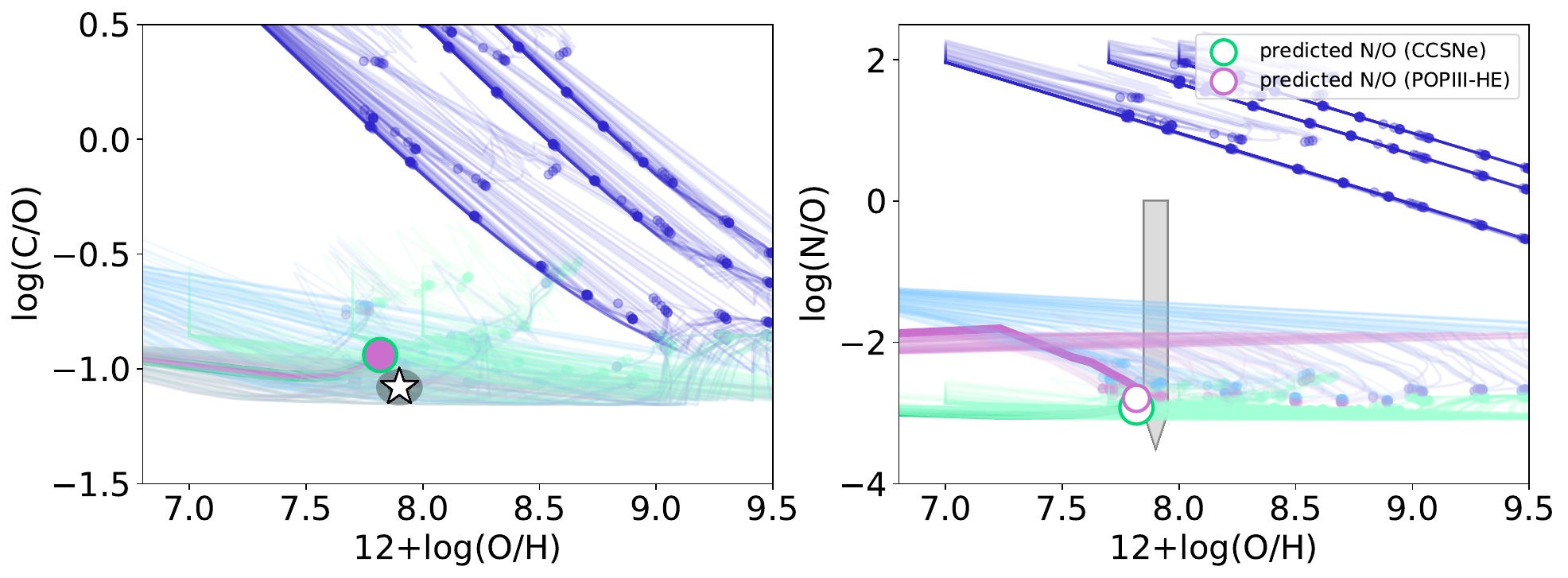}
     \includegraphics[width = 0.9\textwidth]{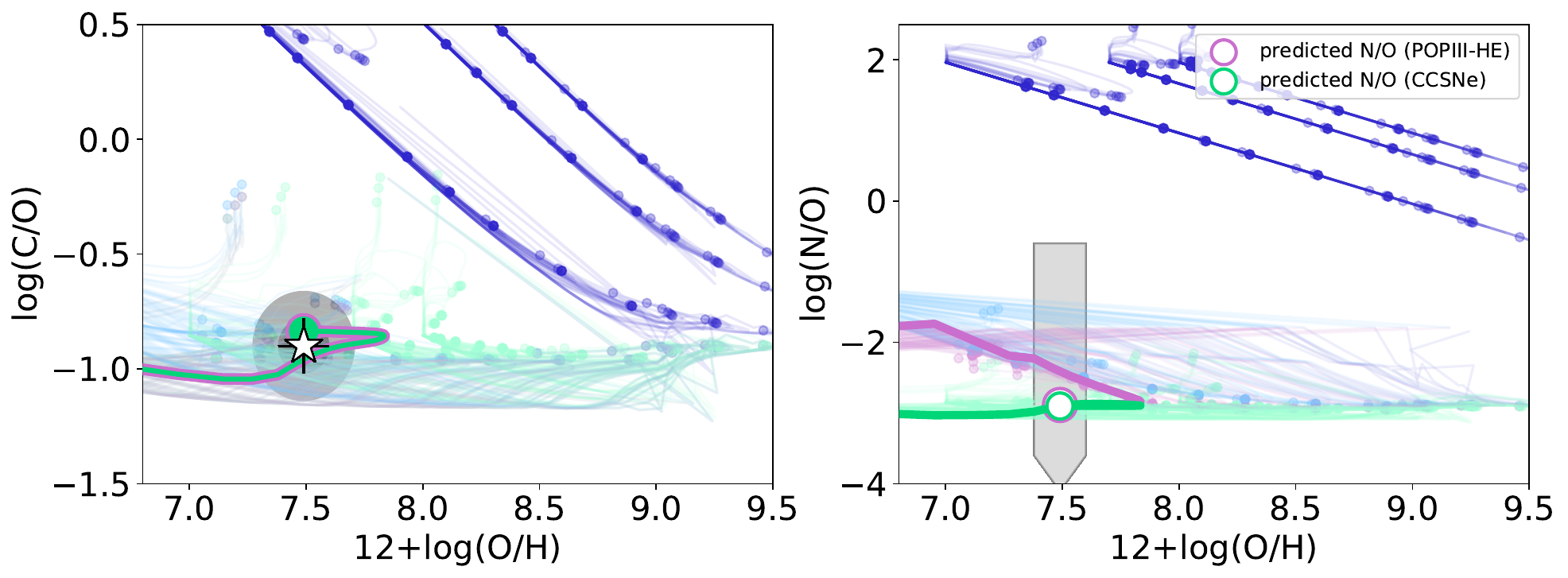}
    \caption{Same as Figure \ref{fig:CO_NO_CEERS}, but for ERO s40590, MACS1149-JD1, and JADES-GS-z9-0. We show upper (or lower) limits as grey upward (downward) triangular shaped area. We use a beige shaded area in case of missing measurement. %The colored curves are predicted by our models. Different color symbols represent different IGM/CGM pre-enrichment by SMSs (green), CCSNE (violet), low explosion energy Population III stars (red), high explosion energy Population III stars (cyan). The best fit model is shown as a thicker curve. The final computation time of each model is marked with small circles (bigger circle in the case of the best fit model). In the log(N/O) vs. 12 + log(O/H) plane, a white circle marks the final computation time of the best-fit model when it corresponds to a predicted value of log(N/O). Note that, for MACS1149-JD1, two possible pre-enrichment scenarios yields equally good $\chi^{2}$ fits (i.e. CCSNe and POPIII-HE ). They overlap in the log(C/O) vs. 12 + log(O/H) plane, but they are distinguishable in the log(N/O) vs. 12 + log(O/H) plane.
    }
\label{fig:main_res_1}
\end{figure*}

\begin{figure*}[htbp]
     \includegraphics[width = 0.9\textwidth]%{plot_main_Curti24_new_DET_apr26.pdf}
     {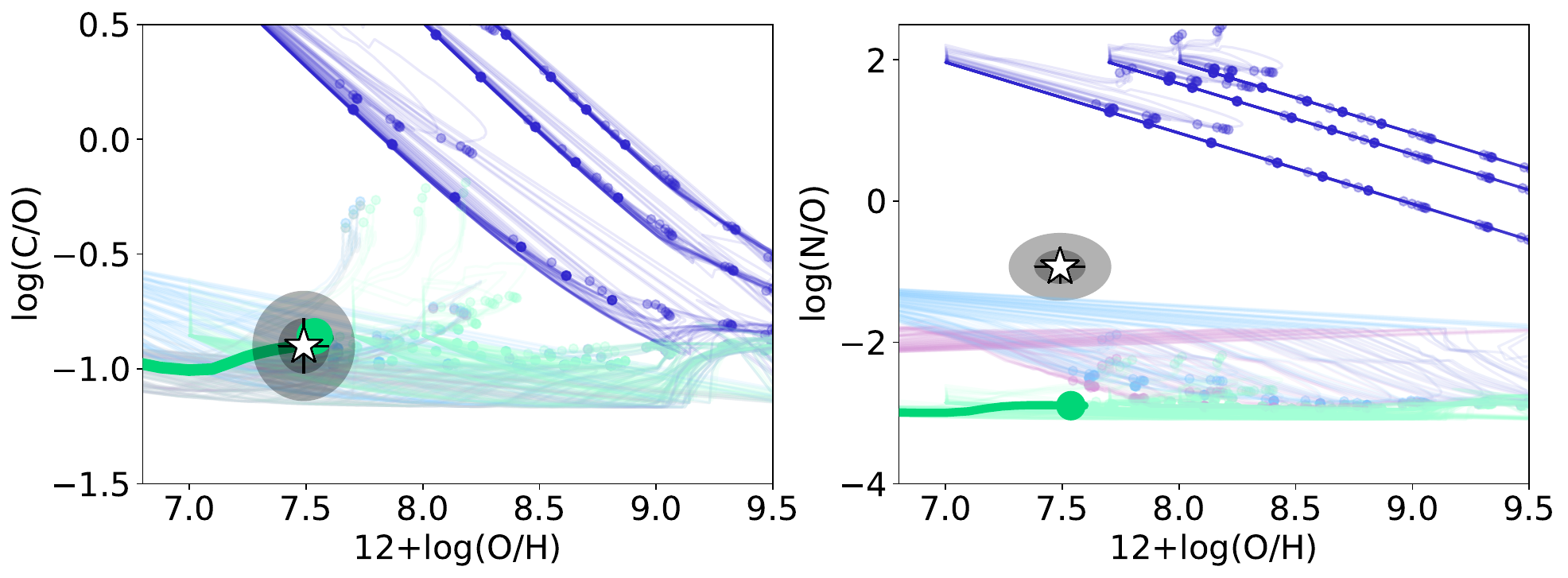}
      \includegraphics[width = 0.9\textwidth]%{plot_main_Curti24_new_oct25_WR.pdf}
      {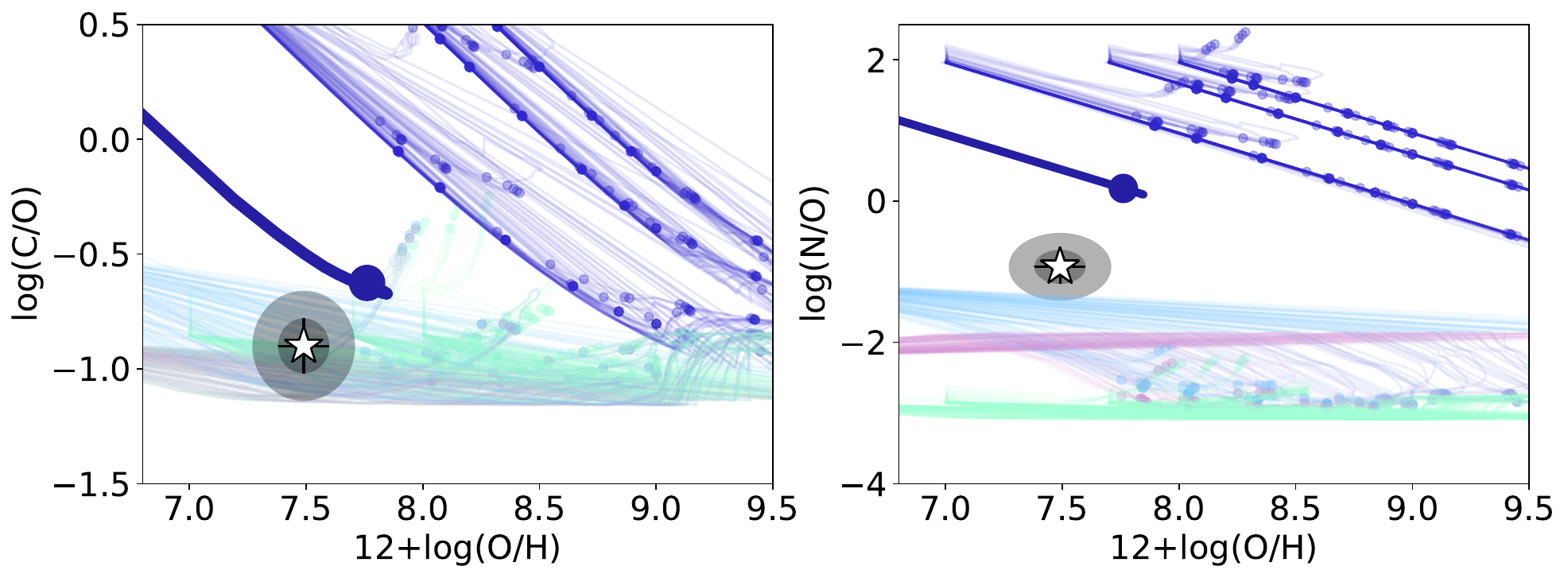}

%old figures:
%plot_main_Curti24_new_oct25_WR.pdf
%{plot_main_Curti24_new_oct25_DET.pdf}
\caption{Same as Figure \ref{fig:CO_NO_CEERS} and \ref{fig:main_res_1}, but for JADES-GS-z9-0 when the $\chi^2$ minimization is conducted on the N/O detection (top panel). In the bottom panel we test the possibility that JADES-GS-z9-0 might contain WR stars. Specifically, we show the models obtained by assuming that stars with mass $\gtrsim40~\rm M_{\odot}$ evolve as WR stars. We find that even when WR stars are assumed, the N/O ratio is underestimated.}
\label{fig:curti24_2}
\end{figure*}

\begin{figure*}[!htpb]
   
    \includegraphics[width = 0.9\textwidth]{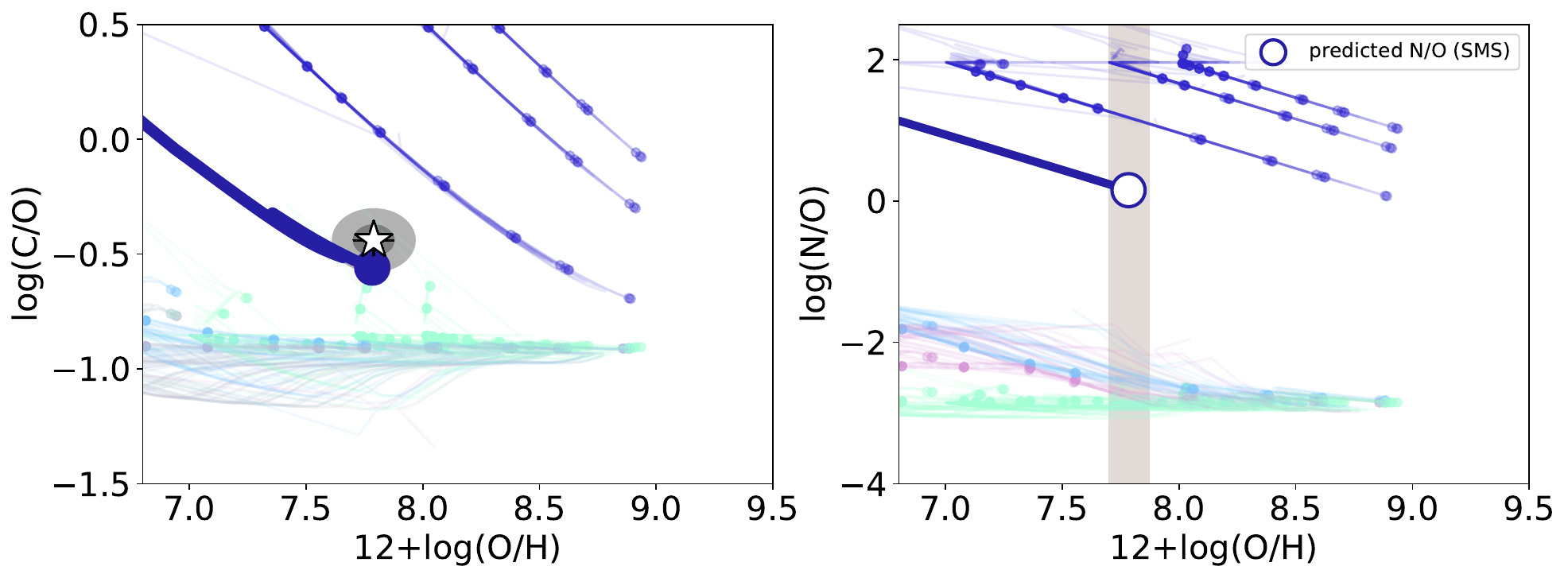}
    \includegraphics[width = 0.9\textwidth]{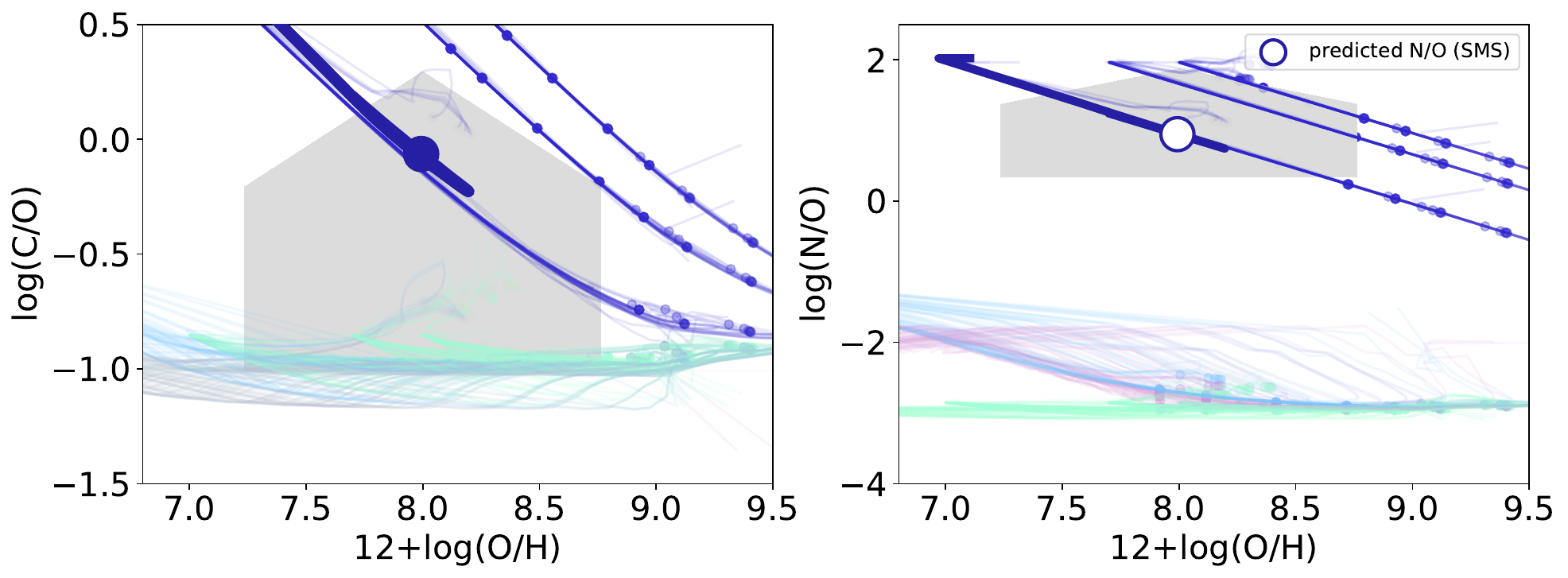}
    \caption{Same as Figures \ref{fig:CO_NO_CEERS} and  \ref{fig:main_res_1}, but for galaxies  MACS0647–JD and GN-z11.}
\label{fig:main_res_2}
\end{figure*}

\begin{figure*}[!htbp]
    \includegraphics[width = 0.9\textwidth]{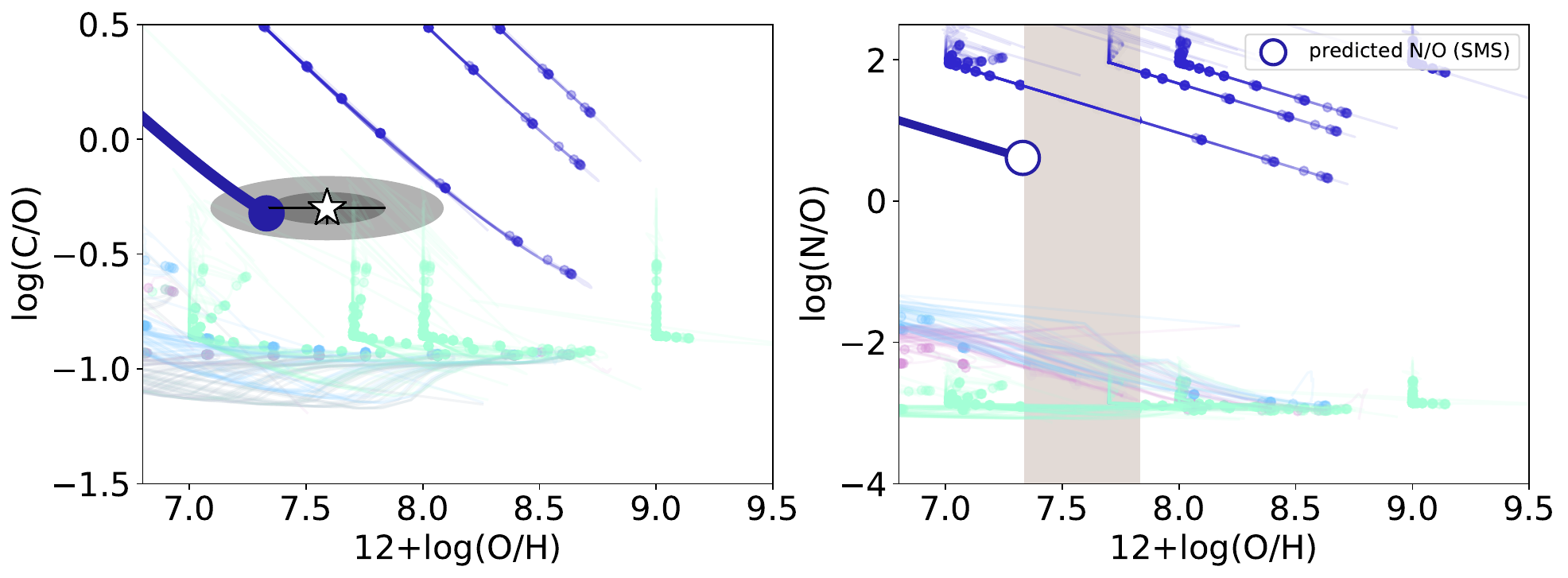}
     \includegraphics[width = 0.9\textwidth]{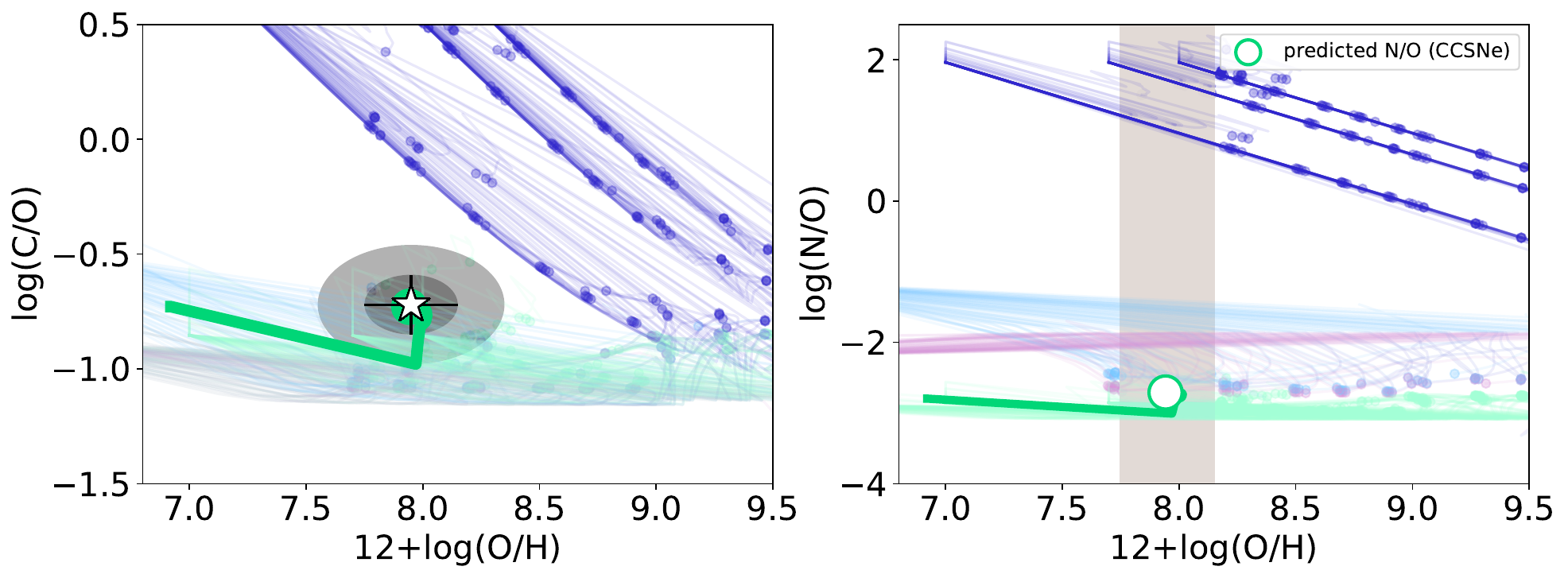}
\caption{Same as Figures \ref{fig:CO_NO_CEERS} and \ref{fig:main_res_1}, but for galaxies GS-z12, and JADES-GS-z14-0.}
\label{fig:main_res_3}
\end{figure*}


\bibliography{bibliography1}{}
\bibliographystyle{aasjournal}


%\listofchan
\end{document}